\documentclass[prb,superscriptaddress,twocolumn,aps,showpacs,amsmath,amssymb,floatfix]{revtex4-1}

\usepackage{color}
\usepackage{bm}
\usepackage{hyperref}
\usepackage{graphicx}
\usepackage{braket}
\usepackage{MnSymbol}
\usepackage{amssymb}
\hypersetup{colorlinks=true}
\usepackage{dcolumn}
\usepackage{amsmath}
\usepackage{amsfonts}
\usepackage{bm}
\usepackage{color}

\hypersetup{colorlinks,linkcolor=blue,urlcolor=blue,citecolor=blue}

\newcommand{\cyan}[1]{\textcolor{black}{#1}}

\newcommand{\blue}[1]{\textcolor{black}{#1}}
\renewcommand{\v}[1]{\textbf{#1}}
\newcommand{\tr}{\text{tr}}

\begin{document}
\title{Triplet resonating valence bond theory and \cyan{transition metal chalcogenides}}
\author{Elio~J.~K\"onig}
\affiliation{Max-Planck-Institut f\"ur Festk\"orperforschung, 70569 Stuttgart, Germany}
\affiliation{Department of Physics and Astronomy, Center for Materials Theory, Rutgers University, Piscataway, NJ 08854, USA}
\author{Yashar~Komijani}
\affiliation{Department of Physics, University of Cincinnati, Cincinnati, OH 45221, USA}
\affiliation{Department of Physics and Astronomy, Center for Materials Theory, Rutgers University, Piscataway, NJ 08854, USA}
\author{Piers Coleman}
\affiliation{Department of Physics and Astronomy, Center for Materials Theory, Rutgers University, Piscataway, NJ 08854, USA}
\affiliation{Department of Physics, Royal Holloway, University of London, Egham, Surrey TW20 0EX, UK}
\date{\today }

\begin{abstract} 
We develop a quantum spin liquid theory for quantum magnets with
easy-plane ferromagnetic exchange. These strongly entangled
quantum states are obtained by dimer coverings of 2D lattices with
triplet $S = 1, m_z = 0$ bonds, forming a triplet resonating valence bond (tRVB) state. We discuss the conditions and the procedure to transfer well-known results from conventional singlet resonating valence bond theory to tRVB. Additionally, we present mean field theories of Abrikosov fermions on 2D triangular and square lattices, which can be controlled in an appropriate large $N$ limit. We also incorporate the effect of charge doping which stabilizes $p+ip$-wave superconductivity. Beyond the pure theoretical interest, our study may help to resolve contradictory statements on certain transition metal chalcogenides, including 1T-TaS$_2$, as potential tRVB spin-liquids.
\end{abstract}

\maketitle
\section{Introduction} 

\subsection{Resonating Valence Bond theory}

Resonating valence bond (RVB) theory describes prototypical quantum spin liquid (QSL) states which were originally proposed by Anderson~\cite{Anderson1973,Fazekas1974} for the 2D Heisenberg antiferromagnet on a triangular lattice. The frustrated magnetic interactions entangle spins on different sites of the lattice in a pairwise fashion into singlet valence bonds. When the system resonates between a multitude of degenerate bond configurations, the RVB state forms a quantum superposition of a macroscopic number of wave functions. RVB states with short range bonds on two dimensional lattices do not break any of the system's symmetries and satisfy modern criteria~\cite{SavaryBalents2016,KnolleMoessner2019} for a quantum spin liquid (QSL) as a highly entangled state with topological order.\,\cite{KivelsonSethna1987} While it is now known that the ground state of the nearest neighbor Heisenberg antiferromagnet on the triangular lattice is not a QSL state, numerical studies suggest that QSL behavior can be stabilized by next nearest neighbor interactions.\,\cite{ZhuWhite2015,HuShen2015,IqbalBecca2016} Short range RVB states are furthermore known to be the exact ground state of dimer models~\cite{RokhsarKivelson1988} and even of appropriately designed SU(2) invariant spin Hamiltonians~\cite{CanoFendley2010} with $n$-spin interactions (including $n > 2$).

\begin{figure}
\includegraphics[width = .45\textwidth]{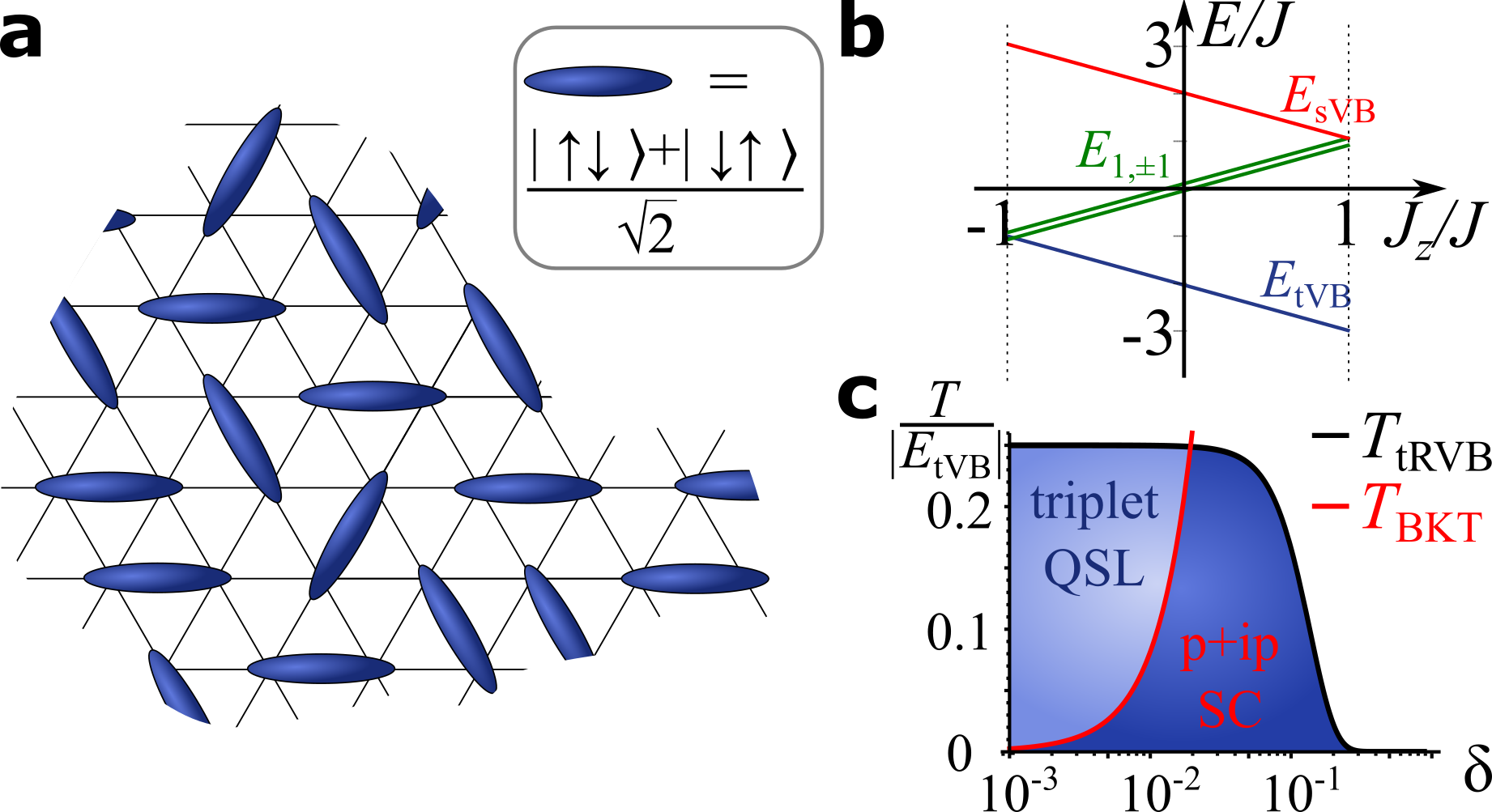}
\caption{{\bf a} Illustration of a dimer covering of the triangular lattice with nearest neighbor triplet bonds (see inset for definition of a dimer). The tRVB ground state is a superposition of such coverings and a triplet quantum spin liquid (QSL). {\bf b} Energy levels of a pair of spins with anisotropic interaction as in Eq.~\eqref{eq:HtVB}, illustrating that the triplet valence bond (tVB) is lowest in energy. {\bf c} Finite temperature mean field phase diagram for the model introduced in Eq.~\eqref{eq:H0} as a function of temperature and hole doping $\delta$. A $p+ip$ superconductor (SC) appears by doping the QSL. For this plot, we used $\vert E_{\rm tVB} \vert/t = 0.1$.}
\label{fig:MainPic}
\end{figure}

	In addition to its progenitorial role in the study of QSLs, RVB theory is also quintessential for superconductivity beyond the BCS paradigm. Specifically in the context of cuprate superconductors, preentangled singlet bonds are believed to constitute a pair condensate which turns into a \textit{bona fide} superconductor as doping liberates the charge degrees of freedom from a correlated Mott insulator.\,\cite{Anderson1987} As a major advantage over other approaches, RVB theory~\cite{KotliarLiu1988,KivelsonEmery1995} and related gauge theories~\cite{Sachdev2018,KoenigTsvelik2019} naturally account for pseudogap phenomena~\cite{ProustTaillefer2019} in an elegant and economical fashion. 

\subsection{\cyan{Spin liquid candidates with ferromagnetic correlations}}

The experimental search for QSL materials has lately enjoyed increased
attention. \cyan{Here, we concentrate on the proposal that the layered
van-der-Waals material 1T-TaS$_2$ and related compounds 1T-NbSe$_2$ and 1T-TaSe$_2$, may form a 2D QSL on a triangular lattice\cite{LawLee2017,KratochvilovaPark2017}, while the 2D ferromagnets CrXTe$_3$ with X=Si,Ge may form 2D QSL on a honeycomb lattice about the T$_{\rm Curie}$.}

When 1T-TaS$_2$ is cooled below 200K, a peculiar CDW order rearranges the atoms of a plane into a triangular superlattice where 13 Ta atoms per unit cell form a star-of-David structure (for a recent STM visualization, see e.g.~\onlinecite{QiaoLiu2017}). As each of the Ta atoms contributes one 5d electron, twelve electrons fill 6 emergent bands and leave an emergent Hubbard model with one electron per supercell behind. Mott localization is generally accepted in view of disorder-dependent activated behavior below~$T \sim 50K$~\cite{FazekasTosatti1979,RibakKanigel2017,MuruyamaMatsuda2020} and the observation of Hubbard bands.\,\cite{QiaoLiu2017,Lutsyk2018} A transition displaying magnetic ordering is ruled out down to temperatures as low as 20mK by $\mu$SR and susceptibility measurements.\,\cite{RibakKanigel2017,KratochvilovaPark2017,Klanjvsek2017}
The ground-state is a paramagnet with a substantial, temperature
independent  magnetic susceptibility $\chi $, 
with a small
Curie-Weiss upturn, likely the result of
disorder~\cite{RibakKanigel2017,KratochvilovaPark2017}.
{These observations raise the fascinating
possibility that this system forms a QSL}, 
a paramagnetic spin state with a characteristic temperature (set by
the spin-spin interaction energy $J$)\cite{LawLee2017}.
{Support for this interpretation derives from the recent}
observation of a finite 
Sommerfeld coefficient $\gamma= C_{V}/T$  in the specific heat 
~\cite{RibakKanigel2017,MuruyamaMatsuda2020} and a 
linear temperature dependence in the thermal
conductivity\,\cite{MuruyamaMatsuda2020}, features that are
consistent with the formation of a QSL with a spinon-Fermi surface. Moreover, the
Wilson ratio $\chi /\gamma$ between the susceptibility and Sommerfeld
coefficient lies in a range that is compatible with weakly interacting spinons. A similar phenomenology and evidence for QSL behavior was recently observed in tunneling experiments on monolayer 1T-TaSe$_2$.~\cite{RuanCrommie2021}

Despite this interesting experimental development, we are not aware of
a microscopic theory which explains the 
large spin-interaction scale $J$ required to account for the
temperature independent susceptibility. Both
a heuristic approach of decoupled star-of-David
clusters~\cite{HeLee2018} and density functional
theory~\cite{YuZou2017} suggest an in-plane hopping strength $t \sim 10 meV$, which taken together with a realistic $U \sim 200 meV$ implies an \textit{antiferromagnetic} superexchange interaction $J \sim {+}5 K$.
{Moreover}, several ab-initio studies of 1T-TaS$_2$ and phenomenologically similar monolayer 1T-NbSe$_2$, 1T-TaSe$_2$ predict a \textit{ferromagnetic} ground state~\cite{YuZou2017,PasquierYazyev2018,Calandra2018,ChenCrommie2020} with exchange constant $J \sim {-}5 K$ and a few percent anisotropy which favors alignment in the basal plane.\,\cite{PasquierYazyev2018,FootnotePasquierYazyev2018} Leaving the conundrum of the smallness of $\vert J \vert $ to future studies, the apparent controversy on the sign of $J$ motivates us to pose the fundamental and fascinating question, whether a quantum spin-liquid {can exist in presence of  ferromagnetic interactions}. 
On the basis of an affirmative answer, we further study the effect of doping away from the Mott limit and discover a time-reversal symmetry breaking (TRSB) $p+ip$-wave superconductor. This is particularly interesting in light of the recent discovery of TRSB superconductivity in 4H-TaS$_2$ (in which monolayers of 1T-TaS$_2$ are alternatingly stacked with metallic 1H-TaS$_2$ monolayers).\,\cite{RibakKanigel2020}

\cyan{Another potential candidate for a tRVB state is CrXTe$_3$ with X=Si, Ge, previously referred to as 2D ferromagnetic semiconductors. These are layered system with Cr$^{3+}$ ions on a honeycomb lattice \cite{}. In the case of CrSiTe$_3$, transport measurements indicate a thermally activated mechanism with an indirect bulk gap of around 0.4eV, consistent with optical measurement. There is a bulk Curie temperature of around 33K, often accompanied by a structural transition,\,\cite{Ron2019} which is enhanced to 80K in the case of monolayer.\,\cite{Lin2016,Liu2016} The origin of the insulating phase at higher temperature is noteworthy: The local structure of the material splits the d-orbital whose orbitals are half-filled at the Fermi level. The narrow bandwidth of d-orbital strongly enhances the correlations between electrons,\,\cite{Zhang2019} leading to a Mott origin in contrast to the \emph{ab initio} calculations.\,\cite{Siberchicot1996} The fact that the Curie temperate is less than the gap indicates that this is a local-moment system with a saturated magnetic moment of 3$\mu_B$/Cr. The ferromagentism between Cr electrons is induced by the super-exchange via Te sites \cite{Zhang2019} which dominates over a direct anti-ferromagnetic super-exchange. Nonetheless, there are anti-ferromagnetic inter-layer correlations, which seem to be responsible for reducing the Curie temperature in the bulk compared to the monolayer. While there is a slight Ising anisotropy in the transition temperature, significant short-range in-plane magnetic correlations persists all the way up to 150K.\, \cite{Williams2015} Under pressure the system becomes metallic and, after a structural transition at P$\sim$9GPa, superconducting with a Tc of around 4K \cite{Cai2020} which is relatively independent of pressure up to around 50GPa. Based on this, we propose that this material is a potential candidate for the tRVB physics and the associated superconductivity upon doping which is discussed in this paper.}

\subsection{RVB theory vs. triplet RVB theory}

In the past, QSL theories with \textit{Ising} interactions (e.g. the
Kitaev model~\cite{Kitaev2006}) were discussed both for ferromagnetic
and antiferromagnetic interactions. {Here, we develop the concept  }
QSLs with ferromagnetic \textit{easy plane} interactions, leading to
the concept of triplet resonating valence bonds (tRVB). This idea
was recently proposed to account for the observation of  strange metal behavior near a ferromagnetic quantum critical point {in the heavy-fermion material CeRh$_6$Ge$_4$}.\,\cite{ShenYuan2020} {It was subsequently proposed that such a tRVB quantum material can be a parent state for triplet superconductivity in Hund's metals, and in particular in Iron-based superconductors}.\,\cite{ColemanKoenig2020,DrouinColeman2021} As we will now explain, the underlying principles of tRVB theory parallel 
Anderson's RVB theory. {The basic building block of the RVB
theory, 
is a singlet valence bond (sVB)}, {formed between two spins at sites
$i$ and $j$}
\begin{equation}
\ket{[i,j]} = \frac{\ket{\uparrow_{i}}\ket{\downarrow_{j}}-
\ket{\downarrow_{i}}\ket{\uparrow}_{j}}
{\sqrt{2}},
 \end{equation}
where the notation $[i,j]=-[j,i]$ is used to reflect the antisymmetry
under spin exchange. This state
{is the ground state of a two-site Hamiltonian with
antiferromagnetic spin-spin interaction }
\begin{equation}\label{}
H_{\rm sVB} = J \vec{\sigma}_{i}\cdot \vec{\sigma}_{j}.
\end{equation}
Here, $\sigma_{x,y,z}$ are Pauli Matrices and $\hbar = 1$ throughout the paper.
In a lattice of spins that interact antiferromagnetically, such bonds
can develop between any pairs of spin, forming the state 
\begin{equation}\label{}
\vert P_{s}\rangle  = \prod_{[i,j]\in P} \ket{[i,j]}
\end{equation}
where P is a particular choice of pairs of spins. In a lattice,
competition between antiferromagnetic spin interactions cause such a
state to resonate between different configurations $\vert P_{s}\rangle
$, forming a quantum-mechanical admixture of all such states, a resonating valence
bond state 
\begin{equation}\label{}
|\hbox{RVB}\rangle  = \sum_{P}A_{P}\vert P_{s}\rangle.
\end{equation}
The simplest example of such a state is the short-range RVB state,
formed from the 
quantum superposition of all possible coverings with nearest neighbor sVB dimers.

By contrast, the basic building block for tRVB theory is a triplet valence bond (tVB)
\begin{equation}
\ket{(i,j)} = \frac{\ket{\uparrow_{i}}\ket{\downarrow_{j}}+
\ket{\downarrow_{i}}\ket{\uparrow}_{j}}
{\sqrt{2}}.
 \end{equation}
This entangled state is the ground state of a two-site Hamiltonian with easy plane ferromagnetic spin-spin interaction 
\begin{equation}\label{eq:HtVB}
H_{\rm tVB} = - J [\sigma_{x,i} \sigma_{x,j} + \sigma_{y,i}\sigma_{y,j}] + J_z \sigma_{z,i}\sigma_{z,j},
\end{equation}
with $J_z > -J, J>0$, Fig.~\ref{fig:MainPic} {\bf b}.
{In analogy to RVB, the tRVB state on a lattice is the
macroscopic superposition}
\begin{equation}\label{}
|\hbox{tRVB}\rangle  = \sum_{P}A_{P}\vert P_{t}\rangle
\end{equation}
of states 
\begin{equation}\label{}
\vert P_{t}\rangle  = \prod_{(i,j)\in P} \ket{[i,j]}
\end{equation}
corresponding to a particular tiling of triplet valence bonds. 
{Crucially, } the singlet and triplet valence bonds 
$\ket{\text{sVB}}, \ket{\text{tVB}}$ both form  Bell pairs and
are therefore 
suitable 
for the construction of highly entangled {QSL} states, Fig.~\ref{fig:MainPic} {\bf a}. 

\begin{figure}
\includegraphics[scale=1]{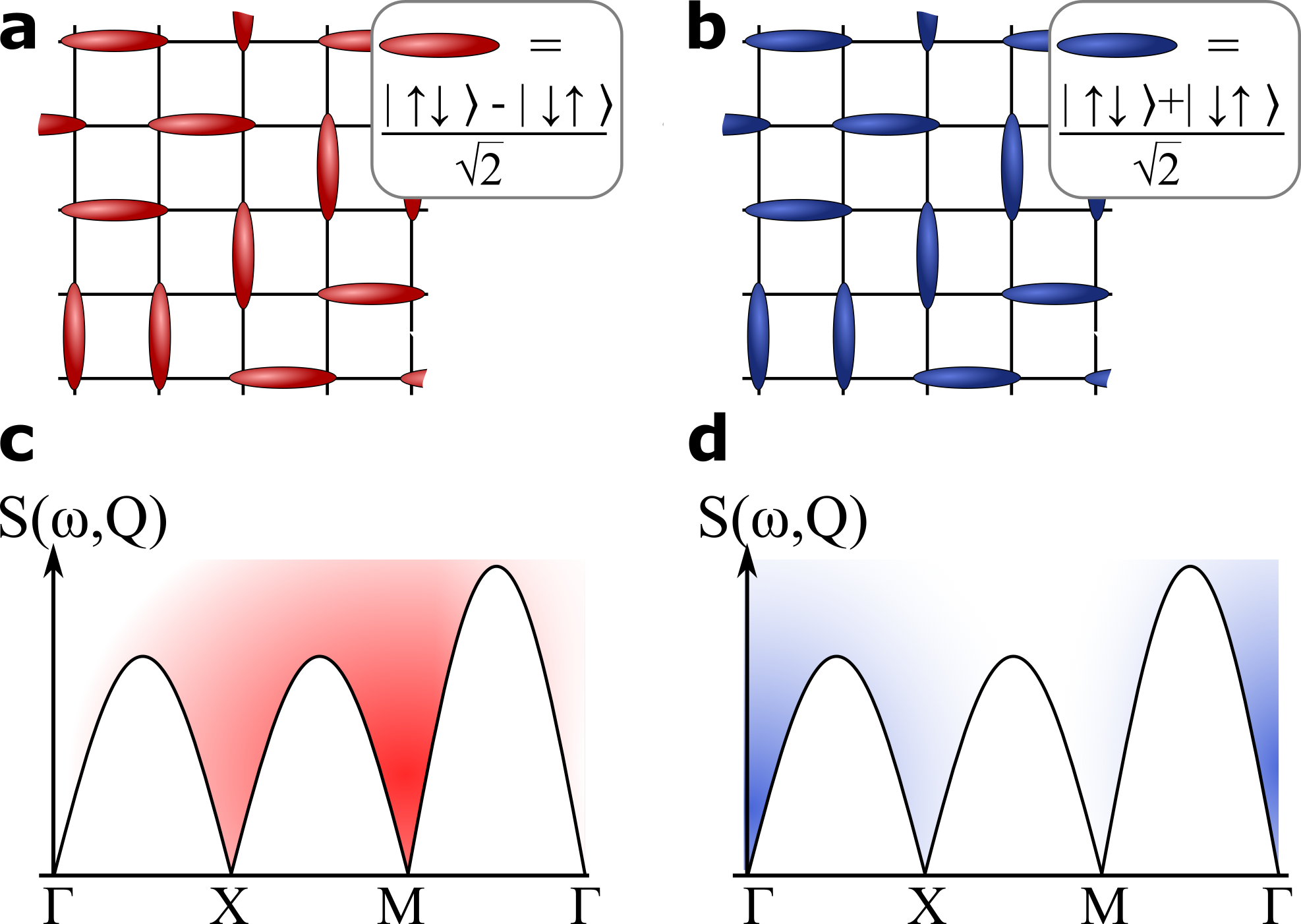}
\caption{The application of a 180$^\circ$ spin rotation on one sublattice, $\vec \sigma_j \rightarrow \sigma_{z,j} \vec \sigma_j \sigma_{z,j}$, transforms a nearest neighbor RVB state on the square lattice (panel a) into a nearest neighbor tRVB state (panel b). By consequence, the dynamical structure factor $S(\omega, \v Q) = \sum_{i}\int dt e^{i \omega t - \v x_i \cdot \v Q} \langle \sigma_{+,i} (t) \sigma_{-,0}(0) \rangle$, $\sigma_\pm = [\sigma_x \pm i \sigma_y]/2$, of the tRVB state (schematic, panel d) is shifted by $(\pi, \pi)$ as compared to the RVB state (panel c).}
\label{fig:RVBtRVB}
\end{figure}

The case where $J_z = J$ is of particular interest, because in this case $H_{\rm
tVB}$ is a unitary transformation of $H_{\rm VB}$ obtained by rotating
the spin at site $j$ through 180$^{\circ}$ about the $z$ axis, i.e 
 $H_{\rm tVB} =  \sigma_{z,j} H_{\rm sVB} 
 \sigma_{z,j}$. Consequently, for a bipartite lattice 
with only intersublattice valence bonds, the tRVB and RVB wave
functions are related by a unitary transformation 
obtained by a 180$^\circ$ rotation of spins on one sublattice. 
Importantly, this implies that nearest neighbor tRVB states on the 2D square lattice are quantum spin-liquids with short range spin correlators, Fig.~\ref{fig:RVBtRVB}, while nearest neighbor tRVB states on the 3D cubic lattice display long range order in $\langle \sigma_{x,i} \sigma_{x,j}\rangle$, $\langle \sigma_{y,i} \sigma_{y,j}\rangle$ and $(-1)^{i+j}\langle \sigma_{z,i}\sigma_{z,j}\rangle$ ($i,j$ are site indices).\,\cite{AlbuquerqueMoessner2012} Moreover, the properties of RVB states which are obtained from quantum dimer models,\,\cite{RokhsarKivelson1988} including the topological ground state degeneracy for the 2D triangular lattice~\cite{MoessnerSondhi2001,FendleySondhi2002,IoselevichFeigelmann2002}, are independent on whether a given dimer represents a singlet or triplet bond. On this basis we conclude that nearest neighbor tRVB theory is a gapped $\mathbb Z_2$ spin liquid on the 2D triangular lattice.

Using these parallels, we here present the first field-theoretical (fractionalized) tRVB theory. 
We fractionalize the electrons into Abrikosov pseudofermions and slave bosons,\,\cite{Coleman1984} and derive a mean field theory of tRVB which is analogous to the mean field approach to the RVB state.\,\cite{AffleckMarston1988,KotliarLiu1988} {Our study particularly focuses on non-bipartite lattices and on the impact of charge doping, because in these two cases the unitary mapping between RVB and tRVB theory breaks down. Most importantly, we thereby develop a formalism for entanglement driven triplet superconductivity:} 
Just as an RVB state may be considered as a Gutzwiller projection (denoted $\hat P_{\rm G}$)
of singlet d-wave superconductive state, i.e. 
\begin{subequations}
\begin{equation}
\ket{\text{RVB}} = {\hat P_{\rm G}} \ket{\text{BCS:} d_{x^2-y^2}},\\
\end{equation}
{the tRVB states considered here are Gutzwiller projected triplet $p_x+ ip_y$ states,}
\begin{equation}\label{eq:tRVBBCS}
\ket{\text{tRVB}} = {\hat P_{\rm G}} \ket{\text{BCS:} p_x + i p_y}.
\end{equation}
\end{subequations}
{At half-filling, both RVB and tRVB theories describe an insulator, yet, upon hole-doping, pre-entangled Cooper pairs get liberated and form a superconducting phase.}

\subsection{{Outline}}

The rest of the paper is structured as follows: In Sec.~\ref{sec:XXZ} we introduce a t-J model for an anisotropic quantum ferromagnet and the formalism of fractionalization. In Sec.~\ref{sec:HomoSol} we present a discussion of homogeneous mean field solutions, both in the Mott limit and upon charge doping. We conclude with an outlook, Sec.~\ref{sec:conclusion}. Three appendices contain details: Appendix~\ref{app:DetailsSpinon} contains technicalities on the free energy expansion, in Appendix~\ref{app:largeN} we demonstrate that an appropriately designed large N limit can stabilize the tRVB mean field solution, while Appendix~\ref{app:BKT} contains details on the Berezinskii-Kosterlitz-Thouless transition establishing a 2D holon superfluid.

\section{XXZ ferromagnet: Model and Formalism}
\label{sec:XXZ}

In this section we introduce model and formalism using the following 2D t-J model with XXZ easy plane ferromagnetic interactions

\begin{eqnarray}
H &=& - t \sum_{\langle i , j \rangle,\sigma } [X_{\sigma 0}^{(i)} X_{0 \sigma}^{(j)} +H.c.] - \mu \sum_{i, \sigma} X_{\sigma \sigma}^{(i)}\notag \\
 &-& J \sum_{\langle i, j \rangle} \lbrace\hat \sigma_x^{(i)}\hat \sigma_x^{(j)}+\hat \sigma_y^{(i)}\hat \sigma_y^{(j)}\rbrace \notag \\
 &+& J_z \sum_{\langle i, j \rangle}  \hat \sigma_z^{(i)}\hat \sigma_z^{(j)} + V \sum_{\langle i, j \rangle} [\sum_\sigma X_{\sigma\sigma}^{(i)}][\sum_\sigma X_{\sigma\sigma}^{(j)}]\rbrace. \label{eq:H0}
\end{eqnarray}

The Hubbard operators {$X_{0 \sigma}^{(i)}=\ket{0}_i\bra{\sigma}_i$, and $X_{\sigma' \sigma}^{(i)}=\ket{\sigma'}_i\bra{\sigma}_i$} satisfy the standard Hubbard superalgebra (see, e.g., Ref~\onlinecite{Colemanbook2015}). The spin operator is $\hat \sigma_\mu = \sum_{\sigma \sigma'} [\sigma_\mu]_{\sigma \sigma'}X_{\sigma \sigma'} $. 

For a given pair of spins, the singlet state $(\ket{\uparrow \downarrow} - \ket{\downarrow \uparrow})/\sqrt{2}$ has energy $E_{\rm sVB}=2J-J_z$, the triplet states $\ket{\uparrow \uparrow}, \ket{\downarrow \downarrow}$ have energy $E_{1,\pm 1}=J_z$ and the $S = 1, m_z = 0$ state $(\ket{\uparrow \downarrow} + \ket{\downarrow \uparrow})/\sqrt{2}$ has energy $E_{\rm tVB}=-2J-J_z$, see Fig.~\ref{fig:MainPic} {\bf b}. The energy scale $E_{\rm tVB}$ of these triplet valence bonds (tVBs) will be crucial throughout the paper, it is manifest that for an easy plane ferromagnet (defined by $J_z>-J$) the ground state of a pair of spins is given by the tVB. 
We consider both triangular and square lattices. 


\subsection{Slave-Boson representation}

We follow the standard slave boson representation~\cite{Colemanbook2015} $X_{\sigma 0}^{(i)} = f^\dagger_{i,\sigma} b_i, X_{\sigma \sigma'}^{(i)} =  f^\dagger_{i,\sigma} f_{i,\sigma'}, X_{00}^{(i)} = b^\dagger_i b_i$ with the local constraint $b^\dagger_i b_i + \sum_\sigma f^\dagger_{i,\sigma} f_{i,\sigma} = 1$. With a slight abuse of language (see details below), we call $f_{i \sigma}$ a spinon and $b_i$ a holon.

The corresponding Hamiltonian is
\begin{eqnarray}
H &=& \sum_{\langle i,j\rangle} \Big[- t (f_{i}^{\dagger} b_i b_j^\dagger f_{j} + H.c.) + J_z(f_i^\dagger \sigma_z f_i)(f^\dagger_j \sigma_z f_j)\notag \\
&&-J \sum_{\mu = x,y}(f^\dagger_i \sigma_\mu f_i ) (f^\dagger_j \sigma_\mu f_j) + V b_i b_i^\dagger b_j b_j^\dagger \Big] \notag \\
&&+ \sum_i[\lambda_i( f_{i}^\dagger f_{i} + b_i^\dagger b_i - 1) - \mu b_i b_i^\dagger ].
\label{eq:HSB}
\end{eqnarray}

We have added a Lagrange multiplier $\lambda_i$ to enforce the local constraint and employ a spinor notation $f_i = (f_{i, \uparrow}, f_{i, \downarrow})^T$. So far, no approximations were made, Eq.~\eqref{eq:H0} was merely rewritten.

As usual, a number of subtleties follow from the prefractionalized construction.
The local nature of the constraints leads to the emergence of a compact U(1) gauge theory (generated by local rotations $ f_{i,\sigma} \rightarrow e^{i \phi_i} f_{i,\sigma}, b_{i} \rightarrow e^{i \phi_i} b_{i}$). Depending on whether the U(1) gauge theory is deconfining or confining, a quantum spin liquid 
with well defined (deconfined) spinons, is or is not realized.\,\cite{KoenigKomijani2020} 
 
It is well known that 2D compact U(1) gauge theories without matter fields are confining due to a proliferations of monopoles.\,\cite{Polyakov1977} While at first sight, this 
suggests that a {truly fractionalized state} 
can not develop from Eq.~\eqref{eq:HSB}, there are essentially three ways to avoid confinement: First, the spinons form a time reversal symmetry broken insulator which leads to the addition of a Chern-Simons term to the gauge theory. Second, the spinons may form a superconductor, and thereby spontaneously ``break'' the symmetry to $\mathbb Z_2$ 
($\mathbb Z_2$ gauge theories are known to allow for deconfinement in 2D.\,\cite{Wegner1971}) In this context, we mention that generically the physical spinons and the $f_{i, \sigma}$ are related but not the same.\,\cite{SenthilFisher2001} Third, when the spinons remain gapless an infinite number of degrees of freedom~\cite{HermeleWen2004} can suppress the proliferation of monopole operators and thereby annihilate the confining effect.

In this paper, which is the first on fractionalization in tRVB theory, we will not study gauge field fluctuations. Instead, we here derive mean field solutions of the spinon Hamiltonian. However, we emphasize that these solutions are superconducting and time reversal symmetry breaking. It is therefore reasonable to expect the possibility of deconfinement in the gauge sector. 

\subsection{Reminder of slave-boson theory for RVB}
\label{sec:RVBreview}

{Before developing the slave-boson theory of tRVB, it appears beneficial to remind the reader about the analogous formalism for conventional RVB.\,\cite{RuckensteinAppel1987,KotliarLiu1988,LeeWen2006}
 We mostly follow Kotliar and Liu~\cite{KotliarLiu1988} and consider the conventional t-J model on the square lattice }

\begin{eqnarray}
H &=& - t \sum_{\langle i , j \rangle,\sigma } [X_{\sigma 0}^{(i)} X_{0 \sigma}^{(j)} +H.c.] - \mu \sum_{i, \sigma} X_{\sigma \sigma}^{(i)}\notag \\
 &+& J \sum_{\langle i, j \rangle} \lbrace\hat {\vec \sigma}^{(i)} \cdot \hat{\vec \sigma}^{(j)} + V \sum_{\langle i, j \rangle} [\sum_\sigma X_{\sigma\sigma}^{(i)}][\sum_\sigma X_{\sigma\sigma}^{(j)}]\rbrace, \label{eq:HAFM}
\end{eqnarray}
which in slave boson formalism may be written as
\begin{eqnarray}
H &=& \sum_{\langle i,j\rangle} \Big[- t (f_{i}^{\dagger} b_i b_j^\dagger f_{j} + H.c.) + J(f_i^\dagger\vec \sigma f_i)\cdot (f^\dagger_j \vec \sigma f_j) \notag \\
 &&+ V b_i b_i^\dagger b_j b_j^\dagger \Big] + \sum_i[\lambda_i( f_{i}^\dagger f_{i} + b_i^\dagger b_i - 1) - \mu b_i b_i^\dagger ].
\label{eq:HAFMSB}
\end{eqnarray}

{The RVB theory summarized in Eq.~\eqref{eq:HAFM} and \eqref{eq:HAFMSB} parallels the tRVB theory exposed in Eq.~\eqref{eq:H0} and Eq.~\eqref{eq:HSB}. Specifically, the interaction part of Eq.~\eqref{eq:H0} at $J_z = J$ is obtained from Eq.~\eqref{eq:HAFM} by applying a unitary transformation $\hat{\vec \sigma} \rightarrow \hat \sigma_z \hat{\vec \sigma} \hat \sigma_z$ on every other site. As a direct consequence, spinon interactions in the slave-boson formulation, Eq.~\eqref{eq:HSB} follow from Eq.~\eqref{eq:HAFMSB} by applying the transformation $f_{j} \rightarrow (\sigma_z)^j f_j$ {on every second site}. However, at $t  \neq 0$, tRVB theory is not related to RVB theory by a mere unitary transformation, because $t (f_i^\dagger b_i b_j^\dagger f_j) \rightarrow t (f_i^\dagger \sigma_z b_i b_j^\dagger f_j)$ leads to a spin-dependent sign of the hopping.}

{One may decouple the spinon-interaction of Eq.~\eqref{eq:HAFMSB} as follows}\,\cite{KotliarLiu1988}
\begin{eqnarray}
H &=& - t \sum_{\langle i,j\rangle,\sigma} [f_{i,\sigma}^{\dagger} b_i b_j^\dagger f_{j,\sigma} + H.c.]\notag \\
&+& \sum_i[\lambda_i(\sum_\sigma f_{i,\sigma}^\dagger f_{i,\sigma} + b_i^\dagger b_i - 1) - \mu b_{i} b_{i}^\dagger ] \notag \\
&-&  \sum_{\langle i,j \rangle} \lbrace( \kappa_{ij} f^\dagger_i f_j + H.c.) - [\Delta_{ij} f_i^T(i \sigma_y) f_j + H.c.]\rbrace\notag \\
&+& \frac{2}{\vert E_{\rm AFM}\vert } \sum_{\langle i,j \rangle}[ \vert \kappa_{ij} \vert^2 + \vert \Delta_{ij} \vert^2] + V \sum_{\langle i,j \rangle} b_ib_i^\dagger b_j b_j^\dagger ,\label{eq:AFMDecoupled}
\end{eqnarray}
{where $E_{\rm AFM} = - 3 J$ is the energy of a single sVB for isotropic antiferromagnetic interactions.} {As first discussed by Affleck, Zou, Hsu and Anderson,~\cite{AffleckAnderson1988} at half-filling, the model displays an $SU(2)$ symmetry in Nambu space.} This is traditionally represented using Nambu spinors $\psi_{i,\sigma} = (f_{i,\sigma},(i \sigma_y)_{\sigma \sigma'} f_{i, \sigma'}^\dagger)^T$, the matrix
\begin{equation}
U_{ij} = \left (\begin{array}{cc}
\kappa_{ij} & \Delta_{ij} \\ 
\Delta_{ij}^* & - \kappa_{ij}^*
\end{array}  \right)
\end{equation}
and the identification
\begin{eqnarray} 
&-& [\kappa_{ij} f^\dagger_i f_j - \Delta_{ij} f_i^T(i \sigma_y) f_j + H.c.]+ 2 \frac{\vert \kappa_{ij} \vert^2 + \vert \Delta_{ij} \vert^2}{\vert E_{\rm AFM}\vert }  \notag \\
&=& - \psi_i^\dagger U_{ij} \psi_j + \frac{\tr[U_{ij}^\dagger U_{ij}]}{\vert E_{\rm AFM}\vert} .
\end{eqnarray}
The symmetry is given by an SU(2) rotation in Nambu space, $\psi_i \rightarrow W_i \psi_i, U_{ij}  \rightarrow W_i U_{ij} W_j$.
{We equivalently represent this symmetry} {in the notation $U_{ij} = i\kappa_{ij}'' + \vec u_{ij} \cdot \vec \tau$}
{as an $SO(3)$ rotation of the vector of order parameters on a given link 
\begin{equation}
\vec u_{ij} = (\Delta_{ij}', \Delta_{ij}'', \kappa_{ij}'),
\end{equation}
 where $\text{Re}[\Delta_{ij}] = \Delta_{ij}', \text{Im}[\Delta_{ij}] = \Delta_{ij}''$ and analogously for $\kappa_{ij}$. }

\begin{figure}
\includegraphics[width = .45\textwidth]{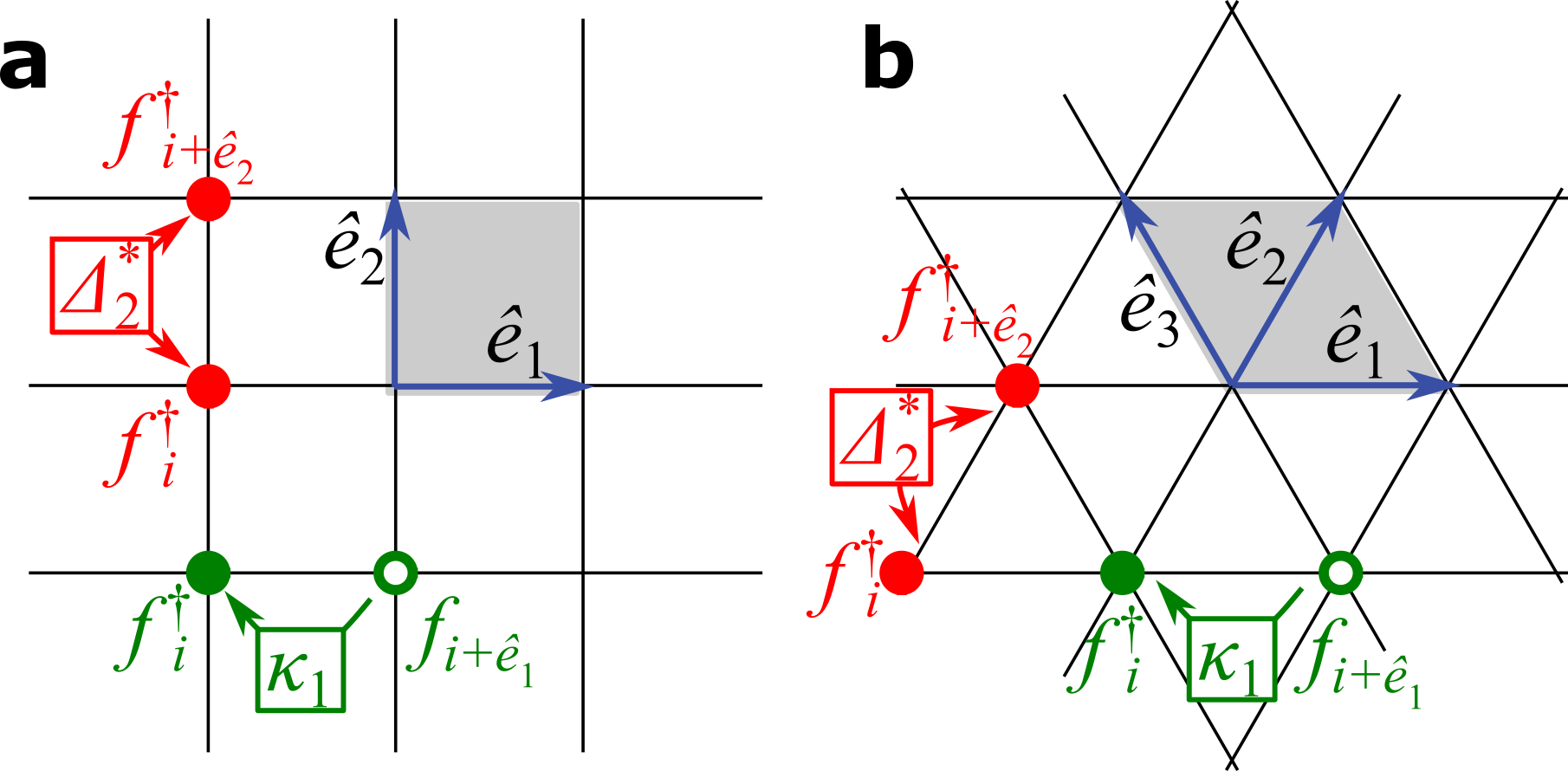}
\caption{Convention of unit vectors $\hat e_l$ ($l = 1, \dots , z/2$) {and illustration of mean field parameters} on {square and} triangular lattices. The homogeneous mean-field solutions discussed in this paper correspond to associating complex hopping $\kappa_l$ and pairing $\Delta_l$ amplitudes to each of the $z/2$ links of each unit cell (which is shaded gray).}
\label{fig:VectorConvention}
\end{figure}

{Kotliar and Liu~\cite{KotliarLiu1988} considered mean field solutions of Eq.~\eqref{eq:AFMDecoupled} under the simplifying assumption of translational invariance, i.e. $\lambda_i =\lambda$, $\kappa_{ij} = \kappa_1$ ($\kappa_{ij} = \kappa_2$) on all horizontal (vertical) links and analogously for $\Delta_{ij}$. Thus, $\kappa_{1,2}$ corresponds to a complex hopping amplitude and $\Delta_{1,2}$ to an intersite pairing gap on a given link, see Fig.~\ref{fig:VectorConvention} {\bf a}. In the absence of doping, the ground state manifold of energetically equivalent solutions is characterized by the condition $\vec u_1 \cdot \vec u_2 = 0$ (i.e. $\vec u_1 \propto (1,0,0)$ and $\vec u_2 \propto (0,1,0)$ and $SO(3)$ rotations theoreof). However, doping adds a linear symmetry breaking field to the free energy, $\delta F_{\rm RVB} \propto - \delta (0,0,1) \cdot [\vec u_1 + \vec u_2]$, such that the manifold of mean field solutions is reduced to d-wave superconducting solutions, i.e. $\vec u_1 \propto (1,0,1), \vec u_2 \propto (-1,0,1)$ or equivalent solutions obtained by $O(2)$ rotations along the axis $(0,0,1)$.}

{Finally, a physical superconductor is reached when both the spinons and the bosons form a superfluid.}

\subsection{Hubbard-Stratonovich decoupling}

{After having reviewed conventional RVB theory, we return to the model of interest in the tRVB context, Eqs.~\eqref{eq:H0}, \eqref{eq:HSB}.} We decouple the interaction in the two channels of strongest nearest neighbor attraction 
\begin{eqnarray}
H &=& - t \sum_{\langle i,j\rangle,\sigma} [f_{i,\sigma}^{\dagger} b_i b_j^\dagger f_{j,\sigma} + H.c.]\notag \\
&+& \sum_i[\lambda_i(\sum_\sigma f_{i,\sigma}^\dagger f_{i,\sigma} + b_i^\dagger b_i - 1) - \mu b_{i} b_{i}^\dagger ] \notag \\
&-&  \sum_{\langle i,j \rangle} \lbrace( \kappa_{ij} f^\dagger_i \sigma_z f_j + H.c.) + [\Delta_{ij} f_i^T(i \sigma_y) \sigma_z f_j + H.c.]\rbrace\notag \\
&+& \frac{2}{\vert E_{\rm tVB}\vert } \sum_{\langle i,j \rangle}[ \vert \kappa_{ij} \vert^2 + \vert \Delta_{ij} \vert^2] + V \sum_{\langle i,j \rangle} b_ib_i^\dagger b_j b_j^\dagger . \label{eq:Decoupled}
\end{eqnarray}

{The $\sigma_z$ matrices in the third line appear for triangular and square lattices alike, in the latter case they directly follow from Eq.~\eqref{eq:AFMDecoupled} by virtue of the transformation $f_j \rightarrow (\sigma_z)^j f_j$.}
The logic for decoupling {particle-hole and particle-particle} channels simultaneously is motivated as follows: In the field integral, we can discriminate particle-particle and particle hole channel from the structure in frequency space: $f^\dagger_{\epsilon_1,\sigma_1}f_{\epsilon_2,\sigma_2}f^\dagger_{\epsilon_3,\sigma_3}f_{\epsilon_4,\sigma_4} \delta_{\epsilon_1 + \epsilon_3 - \epsilon_2 - \epsilon_4}$ has three different channels, according to which of the frequencies $\epsilon_{1,2,3,4}$ are closeby in magnitude. We keep all attractive channels, while the repulsive channels will be dropped.

The procedure of decoupling nearest neighbor channels whilst disregarding onsite {(magnetic)} order parameters is controlled in appropriately designed large N limits (in the present XXZ case of the $Sp(2N)$ group). Extrapolating these calculations to $SU(2) = Sp(2)$ is uncontrolled, even though historically grown.\,\cite{KotliarLiu1988} In particular, the procedure misses all magnetically ordered phases, e.g.~ferromagnetism, to which we compare heuristically in appropriate sections of the main part of this paper.
 
We included a controlled $Sp(2N)$ treatment for the insulating limit in App.~\ref{app:largeN} - this demonstrates that the mean field triplet QSL solutions presented here are the ground state \cyan{of the Hamiltonian (\ref{eq:HSB})} in a well defined limit. For the sake of physical clarity, we consider the formally uncontrolled spin-1/2 system in the main text, and emphasize that we leave the search for spin-1/2 Hamiltonians with rigorous tRVB ground states to the future. 
This strategy parallels the early development of singlet RVB theory,\,\cite{Anderson1973,AffleckMarston1988,KotliarLiu1988} which predated modern, numerically exact, QSL studies, e.g. for antiferromagnetic models on the triangular lattice~\cite{ZhuWhite2015,HuShen2015,IqbalBecca2016,HeLee2018}, by more than three decades.

\section{Homogeneous mean field solutions}
\label{sec:HomoSol}

We here focus on the simplest case of uniform mean field solutions (note that this excludes certain $\pi$ flux solutions~\cite{AffleckMarston1988,RachelThomale2015}) of Eq.~\eqref{eq:Decoupled}. We treat the constraint on average (considering $\lambda$ a constant chemical potential) 
and use a mean field decoupling of the hopping term, by replacing $t \langle b_i b_j^\dagger \rangle \rightarrow t_f$ and $\sum_\sigma t \langle f_i^\dagger f_j \rangle \rightarrow t_b$ \cyan{in Eq.\,(\ref{eq:Decoupled}), i.e.}
\begin{equation}
t {f^\dagger_{i,\sigma}b_i b^\dagger_j f_{j,\sigma}\to t_ff^\dagger_{j,\sigma}f_{i,\sigma}+t_bb_ib_j^\dagger-t_ft_b/t.}
\end{equation}

We will discuss the self-consistency of this replacement in Sec.~\ref{sec:tbtf}, and for the moment we concentrate on the fermionic part of the Hamiltonian.

\subsection{Square lattice}
\label{sec:Square}

{We first discuss the situation on the square lattice, where}
the order parameters are 
$\Delta_{1,2}, \kappa_{1,2}$ (the value on vertical and horizontal links may differ) and $\lambda$ (as determined by the average occupation $1-\delta$). 
The spinon Hamiltonian in Nambu and momentum space is ($\psi_{\v k} =  (  f_{\v k}^T, f^\dagger_{- \v k}(-i \sigma_y )  )^T$)
\begin{subequations}
\begin{align}
H_f &= \frac{1}{2} \sum_{\v k} \psi_{\v k}^\dagger h(\v k) \psi_{\v k} .
\end{align}
It is convenient to express the spinon Hamiltonian $h(\v k)$ using Pauli matrices $\vec \tau$ in Nambu space and the notation $\vec u_l = (\Delta_l', \Delta_l'', \kappa_l'')$ 
\begin{equation}\label{eq:hk}
h(\v k) = \xi_{\v k} \tau_z {- 2 \sum_{l=1}^{z/2} \sigma_z \left [ \kappa_l'\cos(\v k\cdot \hat e_l) - \vec u_l \cdot \vec \tau \sin(\v k\cdot \hat e_l)\right ]}.
\end{equation}
\label{eq:HfSqTr}
\end{subequations}
In this notation, {the previously mentioned}~\cite{AffleckAnderson1988} emergent SU(2) symmetry in Nambu space is manifest in the case $\xi_{\v k} = 0$. 
We introduced $\xi_{\v k} = -2 t_f \sum_{l=1}^{z/2} \cos(\v k \cdot \hat e_l)+\lambda $ as well as the basis vectors $\hat e_1 = (1,0)^T, \hat e_2 = (0,1)^T$, see Fig.~\ref{fig:VectorConvention} a), $z = 4$ is the coordination number. We use the shorthand notation $k_l = \v k \cdot \hat e_l$ in the following.

We {next} evaluate the ground state energy for several trial solutions at $\delta = 0$, see Tab.~\ref{tab:Meanfieldsolutions}, left column.

First, we consider normal state solutions, the simplest of which is $\kappa_1' = \kappa_2' > 0$, $\vec u_l = 0$ (the sign of the hopping can be chosen at will, as spin-up and spin-down spinons have reversed dispersion). {This state displays a spinon Fermi surface, $C_4$ symmetry and $\lambda = 0$ corresponds to half filling.} The fermionic contribution to the ground state energy is $E_f = - \mathcal C \blue{\kappa_l'}$ with $\mathcal C \approx -1.62$. The mean field energy per site is obtain by optimizing $E = E_f + z \blue{\kappa_l'}^2/\vert E_{\rm tVB} \vert$ from which we obtain $E =- \vert E_{\rm tVB} \vert \mathcal C^2/(4z)$, as quoted in Tab.~\ref{tab:Meanfieldsolutions}. (An analogous procedure is used for all of the following states, only the numerical value of $\mathcal C$ changes from case to case). 

As a second normal state solution we consider $\kappa_1'' = \kappa_2'' >0$, while all other parameters $\kappa_l' = \Delta_l' = \Delta_{l}'' = 0$. For the square lattice the total enclosed flux per square vanishes for homogeneous imaginary hopping $\kappa_l'' >0$, thus this solution is {gauge equivalent} to the previously discussed solution with real hopping {(we will see in the next section that such an equivalence does not hold for analogous two states on the triangular lattice)}.

Finally, we consider a solution displaying Dirac nodes: $\vec u_1 = \vert \vec u_1 \vert (1,0,0)^T, \vec u_2 = \vert \vec u_1 \vert (0,1,0)^T$ (all other variational parameters vanish). Amongst the trial solutions, this solution is lowest in energy, see Tab.~\ref{tab:Meanfieldsolutions}, third row. It corresponds to a $p+ip$ superconductor of spinons, {which is nothing but $\ket{\text{BCS:} p_x + i p_y}$ as presented in Eq.~\eqref{eq:tRVBBCS}. Note that the choice of a homogeneous $\lambda$ only respects the constraint of half filled sites \textit{on average}. To impose the Gutzwiller projection on each site amounts to careful integration over gauge field fluctuations which is beyond the scope of this work.}

\subsection{Triangular lattice}
\label{sec:Triangular}
{Next, we repeat the same analysis for the triangular lattice, for which Eq.~\eqref{eq:HfSqTr} holds equally, yet with $z = 6$ (and therefore three sets of order parameter fields $\kappa_{l = 1,2,3}', \vec u_{l = 1,2,3}$) and $\hat e_1 = (1,0)^T, \hat e_2 = (1,\sqrt{3})^T/2, \hat e_3 = (-1,\sqrt{3})^T/2$, see Fig.~\ref{fig:VectorConvention} b). The mean-field solutions are displayed in the right column of Tab.~\ref{tab:Meanfieldsolutions}.}

{First, consider $\kappa'_1 = \kappa'_2 = \kappa'_3 > 0$, for which half-filling implies $\lambda = -0.836 \kappa'_1$. This state displays a Fermi surface and is $C_6$ invariant.} 

As a second normal state solution we consider $\kappa_l'' = -{\vert\kappa_1''\vert} (-1)^l$, while all other parameters $\kappa_l' = \Delta_l' = \Delta_{l}'' = \lambda = 0$. Note that a flux $\pi/2$ ($-\pi/2$) is enclosed in downside (upside) triangles. Contrary to the case of the square lattice, this state is thus not gauge equivalent to the state with real hopping. On the other hand, a variety of superconducting solutions $\kappa_l'  =\kappa_l'' = 0$, $\vert \Delta_{l} \vert \neq 0$ are equivalent by SU(2) {isospin} symmetry. 

Finally, we consider a solution displaying Dirac nodes: $\vec u_1 = \vert \vec u_1 \vert (1,0,0)^T, \vec u_2 = \vert \vec u_1 \vert (0,1,0)^T, \vec u_3 = \vert \vec u_1 \vert (0,0,1)^T$ (all other variational parameters vanish). {As for the square lattice,} this solution is lowest in energy {and} corresponds to a $p+ip$ superconductor of spinons. In the next section, we demonstrate on the basis of a microscopically derived Ginzburg-Landau function, that the Dirac QSL establishes as the dominant instability at finite temperature. 

\begin{table}
\begin{tabular}{|l|c|c|}
\hline 
 &  square lattice & triangular lattice\\ 
\hline 
$\kappa_l' \neq 0$ & \begin{tabular}{c}\includegraphics[scale=.3]{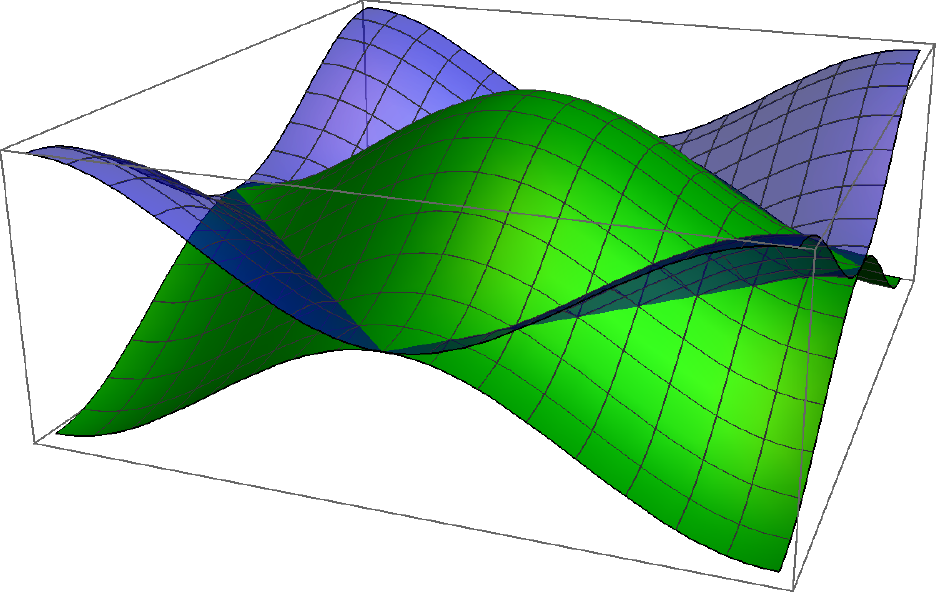}  \\
$E = -0.33 (2J + J_z)$
\end{tabular} &\begin{tabular}{c}\includegraphics[scale=.3]{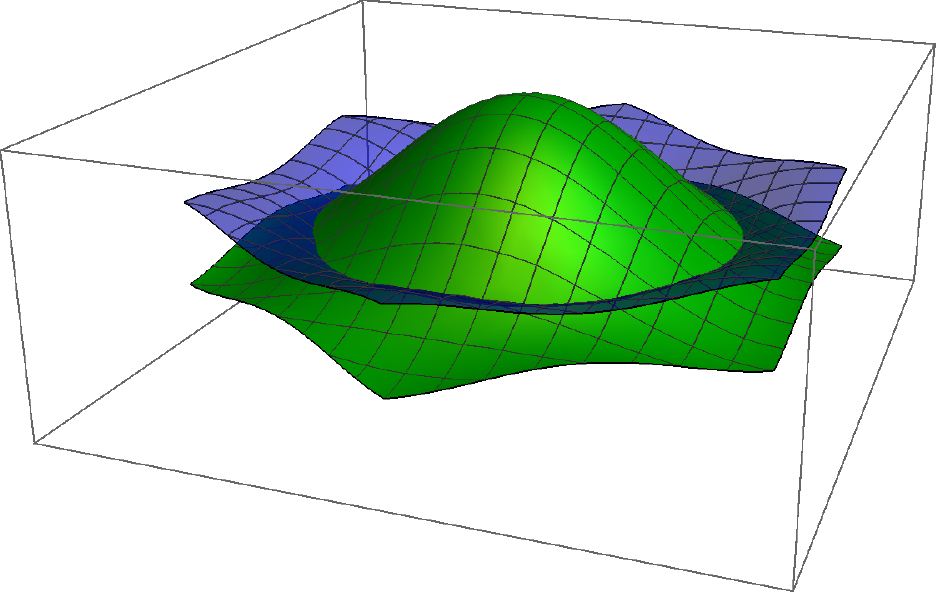}  \\
$E = -0.33 (2J + J_z)$ \end{tabular} \\ 
\hline 
$\kappa_l'' \neq 0$  &\begin{tabular}{c}\includegraphics[scale=.3]{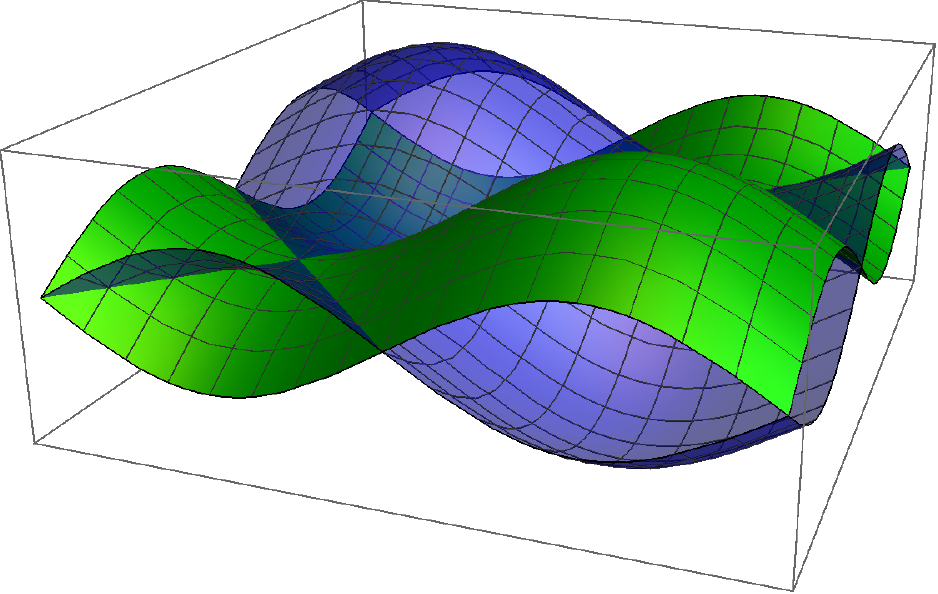}  \\
$E = -0.33 (2J + J_z)$ 
\end{tabular} & \begin{tabular}{c}\includegraphics[scale=.3]{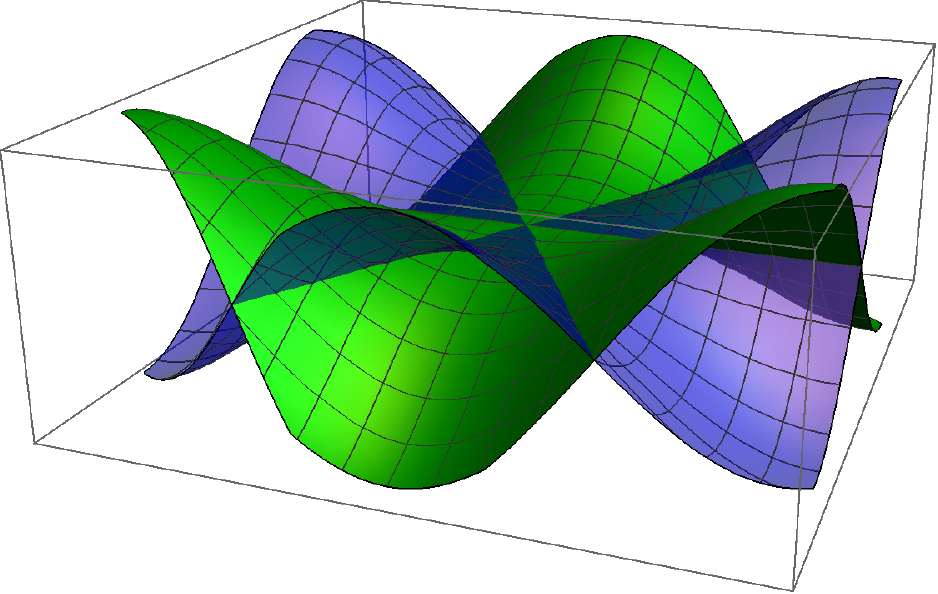}  \\
$E = -0.3 (2J + J_z)$ \end{tabular}\\ 
\hline 
Dirac  &\begin{tabular}{c}\includegraphics[scale=.3]{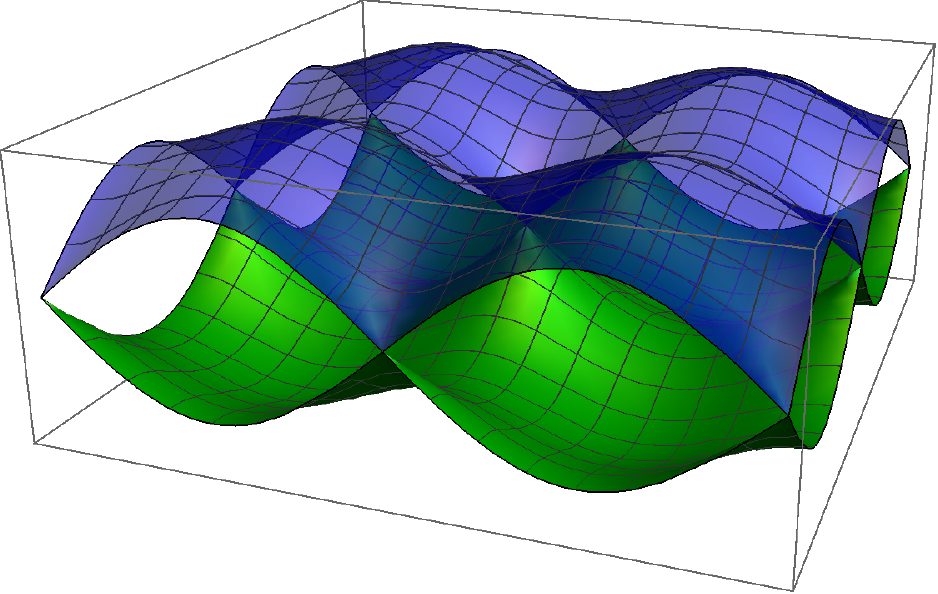}  \\
$E = -0.69 (2J + J_z)$
\end{tabular} &\begin{tabular}{c}\includegraphics[scale=.3]{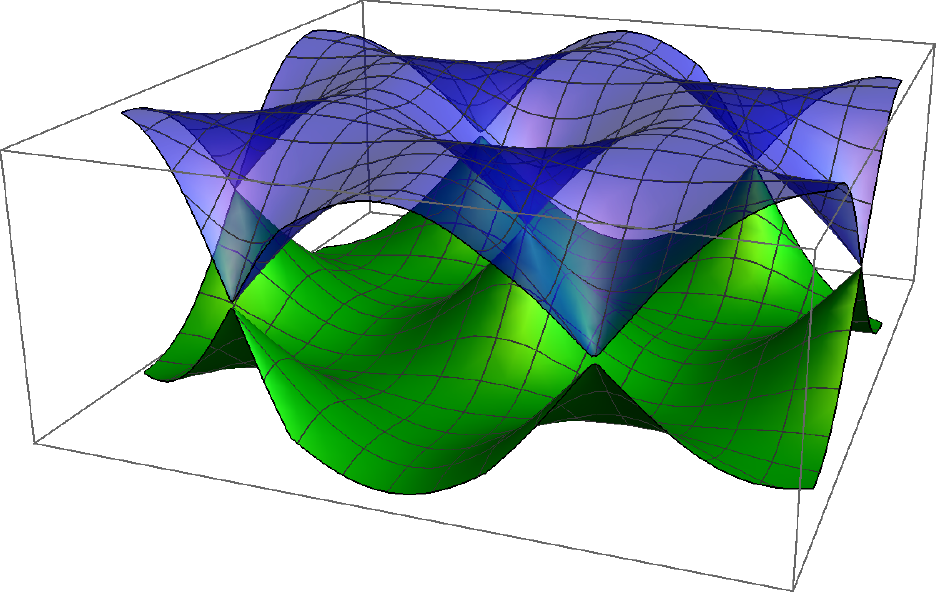}  \\
$E =  -0.47 (2J + J_z)$ \end{tabular} \\
\hline
\end{tabular} 
\caption{Illustration of spinon dispersion in the first Brillouin zone and estimate of the ground state energy (per site) for selected trial mean-field solutions on {square and triangular lattices}, as discussed in Secs.~\ref{sec:Square},\ref{sec:Triangular}. The Dirac solution is $\vec u_1 \perp \vec u_2 \perp \vec u_3$ ($\vec u_1 \perp \vec u_2$) on the triangular (square) lattice. For comparison, the energy of the ferromagnetic in plane solution is $-3J$ ($-2J$) for triangular (square lattice).}
\label{tab:Meanfieldsolutions}
\end{table}

\subsection{Finite temperature transition and doping}
\label{sec:GL}

We here consider the finite temperature transition and the effect of doping on the mean field spinon solution {treating square and triangular lattices in parallel.} This leads to the phase diagram presented in Fig.~\ref{fig:MainPic} {\bf c}. In the calculation we distinguish two regimes: the limit of a degenerate Fermi gas $t_f \gg T$ and the reverse high temperature limit $t_f \ll T$.

We anticipate the result of Sec.~\ref{sec:RealTc} that $t_f = \delta t$ in all relevant regimes and that the degenerate Fermi gas (high temperature classical gas) limit is important for the finite temperature transition at large (small) doping $\delta \gg \vert E_{\rm VB} \vert/t$ ($\delta \ll \vert E_{\rm VB} \vert/t$), see end of this section.

The integration of fermions leads to the following free energy density
\begin{align}
F &= - \frac{T}{2\mathcal A} \sum_n \sum_{\v k} \tr \ln [i \epsilon_n - h({\v k})] + \frac{2}{\vert E_{\rm tVB}\vert } \sum_{l=1}^{z/2}({\kappa'_l}^2 + \vert \vec u_l \vert^2) . \label{eq:FreeEnergy}
\end{align}
Here, $\mathcal A$ is the total number of sites (i.e. the system size). 
The expansion of the fermionic determinant in small order parameter fields leads to a Ginzburg-Landau functional (see Appendix~\ref{app:FreeEnergyExpansion})
\begin{align}
F &= \sum_{l = 1}^{z/2} \left [A_+ \vec u_l^T\vec u_l - A_- \vec u_l^T \left ( \begin{array}{ccc}
1 & 0 & 0 \\ 
0 & 1 & 0 \\ 
0 & 0 & -1
\end{array}\right) \vec u_l  +  C \vert \vec u_l \vert^4 \right ]\notag\\
& + \frac{D}{2} \sum_{l,l'= 1}^{z/2} [ \vec u_l^2 \vec u_{l'}^2 + 2 (\vec u_l \cdot \vec u_{l'})^2 ]. \label{eq:GL}
\end{align}
Here, we disregarded $\kappa'$, which has a subdominant critical temperature in both limits $t_f \ll T$ and $t_f \gg T$. We also emphasize that the constraint field $\lambda$ does not couple linearly to any of the order parameter fields, this is evident from the matrix structure of $h(\v k)$, Eq.~\eqref{eq:hk}.

{It is worthwhile to point out the main difference to conventional RVB theory, namely the absence of a linear symmetry breaking term of the kind $\delta F_{\rm RVB} \propto - (0,0,1) \sum_l \vec u_l$. This is a consequence of the different matrix structure in spin space of the order parameter fields, {i.e. presence of $\sigma_z$ in the decoupled terms in Eq.\,(\ref{eq:Decoupled}). At any finite $\delta$, the isospin SU(2) symmetry is broken by the $A_-$ term.}}

We first qualitatively discuss the mean field solutions assuming $C = 0$, $A_- \ge 0$ and $A_+ \propto T-T_{\rm tRVB}$, where $T_{\rm tRVB}$ is the mean field transition temperature. In the important parameter regime, these assumptions are consistent with the microscopically derived values presented at the end of this section.

The quartic term with a scalar product is crucial in discriminating the lowest energy state and favors $\vec u_l$ which are perpendicular on different bonds $l = 1, \dots, z/2$ of the unit cell. {This reproduces analogous results of singlet RVB theory on the square lattice, which we reviewed in Sec.~\ref{sec:RVBreview}.}
{In contrast to singlet RVB theory, however,} 
the quadratic anisotropy {term $A_-$} favors easy plane solutions in which the superconducting components of $\vec u_l = (\Delta_l', \Delta_l'', \kappa''_l)$ are dominant. More specifically, for the triangular lattice, the mean field order parameter is $\vec u_l \propto ((-1)^l \hat e_l^T, u_z)^T$, where $u_z = {\sqrt{1-3 A_-}}/{\sqrt{4A_-+2}}$ interpolates between the Dirac solution, Tab.~\ref{tab:Meanfieldsolutions} lowest row, and a p-wave superconducting solution with complex nearest neighbor pairing $\Delta_l \propto e^{i 2\pi l/3}$ and $\kappa''_l =0$. In the case of the square lattice, where the coordination number $z$ is smaller, the p-wave superconducting solution is the ground state for any $A_-$. In this case $\Delta_l \propto e^{i \pi l/2}$ and $\kappa''_l =0$.

We proceed with a discussion of microscopic values of the Ginzburg-Landau parameters.
In the high temperature regime $t_f \ll T$, we find $A_- \sim t_f^2/T^3>0$, $C = 0$, $D \sim T^{-3} $ and $A_+ \sim  [T - T_{\rm tRVB} ]/T^2$ with $T_{\rm tRVB} \simeq \vert E_{\rm tVB}\vert/4$. Using $t_f = \delta t$, we find that the regime $t_f \ll T$ is relevant to the finite temperature transition at low doping, $\delta \ll \vert E_{\rm VB} \vert/t$. The emergent SO(3) symmetry at $A_- \propto t_f^2 \rightarrow 0$ ($\vec u_l \rightarrow O \vec u_l$) reflects the SU(2) invariance in Nambu space in Eq.~\eqref{eq:hk}.
This SU(2) symmetry is weakly broken in the regime of weak doping $0<t_f/T \ll 1$ which, as mentioned, favors the superconducting state. 

This tendency is strongly reinforced in the complementary regime of the degenerate electron gas $t_f \gg T$, because $A_+ - A_-$ develops a logarithmic Cooper instability while all other constants remain finite (note that in this case generically $C \neq 0$). In this limit, the mean field transition temperature $T_{\rm tRVB} \sim t_f e^{- \mathcal C t_f /\vert E_{\rm tVB}\vert}$ where we estimate $\mathcal C \simeq 16 \pi \sqrt{3}$ ($\mathcal C \simeq 16 \pi$) for triangular (square) lattice in the continuum limit. Using again $t_f = \delta t$, the degenerate electron gas assumption applies to the finite temperature transition at strong doping $\delta \gg \vert E_{\rm VB}\vert/t$. In Fig.~\ref{fig:MainPic} {\bf c} we present an interpolation of the mean field transition temperature which captures both regimes $\delta \ll \vert E_{\rm tVB}
\vert/t$ and $\delta \gg \vert E_{\rm tVB}\vert/t$.

\begin{figure*}
\includegraphics[scale=1]{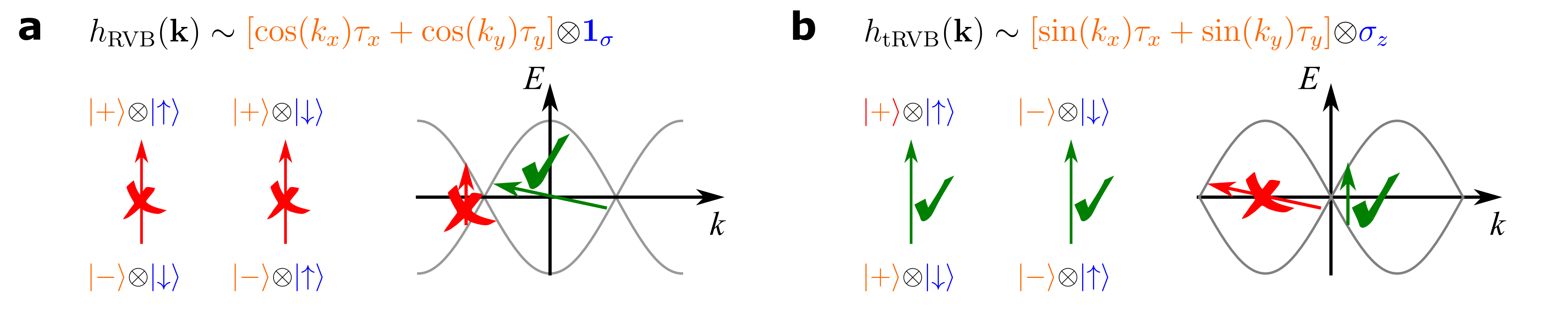}
\caption{Selection rules for the dynamical structure factor for a) the RVB state and b) the tRVB state on the square lattice. For further explanations, see main text.}
\label{fig:SelectionRules}
\end{figure*}

\subsection{Observables}

We {emphasize} that the mean field solutions, which are characterized by $\kappa' = 0$, do not display a finite magnetization $\vec m = T \sum_{n,\v k}\tr[\vec \sigma \mathcal G_{n, \v k}]$, as can be readily seen from the matrix structure of $\mathcal G_{n, \v k}^{-1} = i\epsilon_n - h(\v k)$. At the same time, short range magnetic correlations are key observables and reflected in the dynamical structure factor, Fig.~\ref{fig:RVBtRVB}. We here explain its characteristic features using the selection rules implied by the mean field Hamiltonian of Abrikosov fermions, Fig.~\ref{fig:SelectionRules}. Both RVB and tRVB Hamiltonians are characterized by a direct product of a momentum dependent matrix in Nambu space and a matrix in spin space which is unity in the case of RVB and $\sigma_z$ for tRVB. 
At each momentum $\v k$, we denote the eigenstate with positive (negative) eigenvalue of the matrix in Nambu space by $\ket{+}$ ($\ket{-}$). Hence, at each $\v k$, $\ket{-} \otimes \ket{\uparrow}, \ket{-} \otimes \ket{\downarrow}$ are the negative energy states of $h_{\rm RVB}(\v k)$, and positive energy states are $\ket{+} \otimes \ket{\uparrow}, \ket{+} \otimes \ket{\downarrow}$. 
Thus, the matrix element for vertical spin flip transitions vanishes and the dynamical structure factor is suppressed at zero momentum transfer, $\v Q = 0$, see Fig.~\ref{fig:RVBtRVB} c). 
For the square lattice, $h_{\rm RVB}(\v k) = - h_{\rm RVB} ( \v k + {\v Q}_{\text{N{\'e}el}})$ and spin flip matrix elements at momentum difference ${\v Q}_{\text{N{\'e}el}} = (\pi, \pi)$ are maximal and by consequence the structure factor is dominated by the N{\'e}el wave vector, green arrow in Fig.~\ref{fig:SelectionRules} a). 
In contrast, for $h_{\rm tRVB}(\v k)$, $\ket{-} \otimes \ket{\uparrow}$ and $\ket{+} \otimes \ket{\downarrow}$ are the negative energy states, their spin-flip matrix element with positive energy states $\ket{-} \otimes \ket{\downarrow}$ and $\ket{+} \otimes \ket{\uparrow}$ is finite and vertical transitions are therefore allowed, Fig.~\ref{fig:SelectionRules} b). This leads to a predominantly ferromagnetic spin fluctuation spectrum, Fig.~\ref{fig:RVBtRVB} d).

\subsection{Bosonic Hamiltonian and BKT transition}
\label{sec:RealTc}

So far, we have ignored the slave bosons and simply replaced $t \langle b_i b_j^\dagger \rangle \rightarrow t_f$ in Eq.~\eqref{eq:Decoupled}. In this short section, we reverse the situation and study the bosonic part of Eq.~\eqref{eq:Decoupled} under the replacement $t \sum_\sigma\langle f_{i,\sigma} f_{j,\sigma}^\dagger \rangle \rightarrow t_b$.

We consider the low-density limit, for which the 2D Bose gas is described by the continuum action
\begin{equation}\label{eq:ContBosons}
S = \int d \tau d^2x \bar \psi \left [\partial_\tau - \mu_{\rm eff} - \frac{\nabla^2}{2m} \right ] \psi + \frac{g}{2} \vert \psi \vert^4.
\end{equation}
2D Bose-Einstein condensation is known to be absent in the non-interacting limit and driven by the confinement of topological defects, otherwise. 
The Berezinskii-Kosterlitz-Thouless transition temperature which captures these aspects in the limit $mg \ll 1$ is~\cite{ProkofevSvistunov2001}
\begin{equation}\label{eq:BKT}
T_{\rm BKT} = \frac{2\pi}{m} \frac{n}{\ln\left (\frac{380}{m g}\right)},
\end{equation}
where $n$ is the density in the continuum limit. 

\begin{figure}
\includegraphics[width = .45 \textwidth]{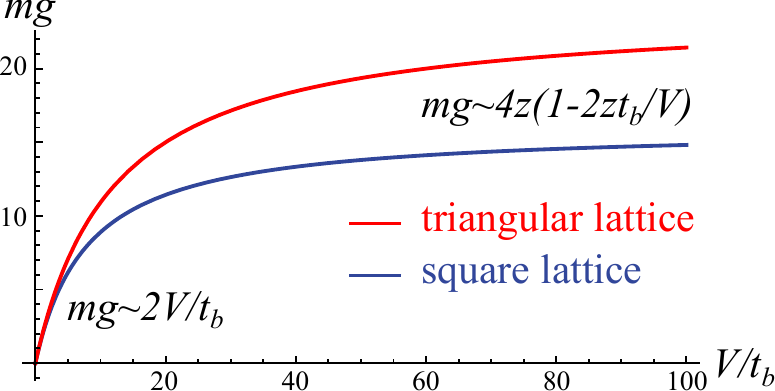}
\caption{The dimensionless parameter $mg$ entering Eqs.~\eqref{eq:ContBosons},\eqref{eq:BKT} as a function of microscopic parameters of the bosonic theory Eq.~\eqref{eq:Decoupled}.}
\label{fig:mg}
\end{figure}

The mapping between the lattice Hamiltonian and the continuum theory {as well as microscopic values for the parameters of {the action} \eqref{eq:ContBosons} are contained in} App.~\ref{app:BKT} {and lead to a non-trivial relationship between $mg$ and $V/t_b$, see Fig.~\ref{fig:mg}}. We exploit this as we push Eq.~\eqref{eq:BKT} to the limit of its applicability at $mg \sim 1$ using 
\begin{equation}\label{eq:TBKTMain}
T_{\rm BKT} = \frac{2\pi z t_b \delta}{\ln[95(1+2 z t_b /V)/z]}
\end{equation}
to interpolate between weak and strong coupling.

\subsection{Self-consistency of hopping amplitudes}
\label{sec:tbtf}
\begin{figure}
\includegraphics[width = .45 \textwidth]{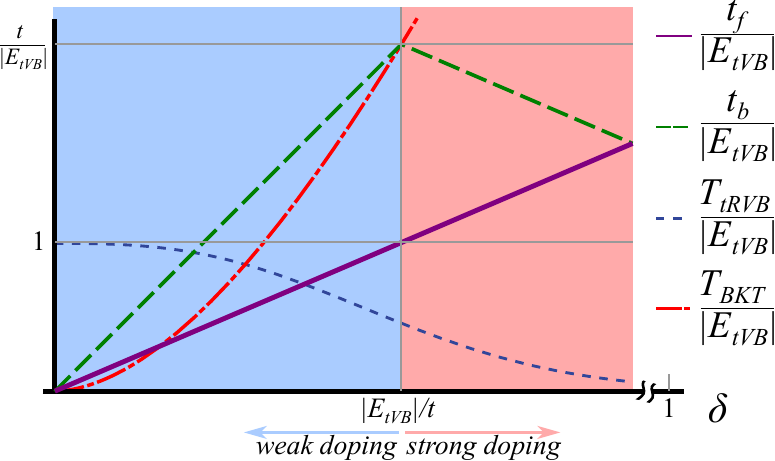}
\caption{{Estimated relationship between internal parameters $t_{b}, t_f$ and external parameters $\delta, t, E_{tVB}$ (note $t \gg \vert E_{tVB}\vert$ is assumed).}}
\label{fig:Regimes}
\end{figure}

 We now {present estimates for} $t_f$ and $t_b$ as a function of external parameters $\delta, E_{\rm tVB}, t$ and first summarize our results: 
 
{For the entire relevant parameter regime, $t_f = \delta t$ while $t_b$ has a different form in weak doping regime  $\delta < \vert E_{\rm tVB} \vert/t \ll 1$ and strong doping regime $ \vert E_{\rm tVB} \vert/t < \delta$. In the weak doping regime, $t_b \sim \delta t^2/\vert E_{\rm tVB} \vert$ and in the strong doping regime $t_b \sim (1 - \delta) t$.}
{For a pictorial summary of these results and associated regimes, see Fig.~\ref{fig:Regimes}}. Formally, the results {for the BKT transition} in the weak doping regime are valid for $T \ll  t_b$, while for {the BCS like transition at} strong doping we require $T\ll t_f$. {Both} $T_{\rm tRVB}$, $T_{\rm BKT}$ satisfy these bounds (we assume $V \sim \vert E_{\rm tVB}\vert$), see also Fig.~\ref{fig:MainPic} {\bf c}.

To derive these results, we first concentrate on the limit $t_f \ll \max_l \vert \vec u_l \vert$. It is important to emphasize that, contrary to usual singlet RVB theory,\,\cite{KotliarLiu1988} magnetic interactions do not contribute to the fermion hopping term proportional to $t_f$, see Eq.~\eqref{eq:hk}. Therefore (assuming nearest neighbors $i$ and $j$),
\begin{align}
t_b &= t \sum_\sigma \langle f_{i, \sigma}^\dagger f_{j, \sigma} \rangle\simeq \frac{2T t}{\mathcal A} \sum_{\v k,n}  \frac{-\xi_{\v k}e^{i \v k \cdot (\v x_i -\v x_j)}}{\epsilon_n^2 + \vert \sum_{l} \vec u_l  \sin(\v k \cdot \hat e_l)\vert^2}
\end{align}
Without going into details, we conclude that $t_b \sim t_f t/\vert E_{\rm tVB} \vert >0$ at temperatures far below $T_{\rm tRVB}$.

On the other hand, in the limit of dominant normal hopping, $t_f \gg \max_l \vert \vec u_l \vert$, one may omit the mean field order parameter in the evaluation of $t_b = t \sum_\sigma \langle f_{i,\sigma}^\dagger f_{j,\sigma} \rangle \simeq t (1- \delta)$. The approximate replacement of the fermionic correlator by the fermion density $1-\delta$ becomes exact in the continuum limit.

Finally, we consider $t_f = t \sum_\sigma \langle  b_{j}^\dagger b_{i} \rangle$. It is obvious that $t_f = \delta t$ in the superfluid, where $\langle b_i \rangle = \sqrt{\delta} e^{i \phi_i}$. We now show that $t_f = \delta t$ also in the normal state and exploit that the relevant regime regards low densities. In this limit the bosonic Hamiltonian can be linearized, $H = \sum_{\v k} b_{\v k}^\dagger \xi_{\v k}^b b_{\v k}$, where $\xi_{\v k}^{b}$ has a minimum at $\v k = 0$ and bandwidth $t_b$. Then
\begin{equation}
t_f \simeq \frac{t}{\mathcal A} \sum_{\v k} e^{i \v k \cdot (\v x_i -\v x_j)} n_{\rm BE}(\xi_{\v k}^b) \simeq \delta t.
\end{equation}
Here $n_{\rm BE}(\xi_{\v k}^b)$ is the Bose-Einstein distribution. At the second asymptotic equality sign, we have used the continuum limit (expansion about $\v k = 0$), which is justified for temperatures $T \ll  t_b$. 

We conclude with a remark that the discrimination of two regimes $t_f \ll  \max_l \vert \vec u_l \vert$, $t_f \gg  \max_l \vert \vec u_l \vert$ at temperatures far below $T_{\rm tRVB}$ is equivalent to $\delta t \ll \vert E_{\rm tVB} \vert, \delta t \gg \vert E_{\rm tVB} \vert$, respectively.  

\section{Conclusion}
\label{sec:conclusion}

In summary, we have studied tRVB states on 2D triangular and square lattices using an Abrikosov fermion mean field treatment of an anisotropic nearest neighbor ferromagnetic Heisenberg model. We found a gapless Dirac spin liquid for either lattice. We furthermore studied the effect of doping away from the Mott insulator limit and thereby discovered a novel mechanism for the appearance of $p+ip$ triplet superconductivity. 

We conclude with an outlook. On the abstract theoretical side, a question about the stability of the tRVB mean field solutions arises for spin 1/2 Hamiltonians. To appreciate the exigency of this question for the present ferromagnetic model Eq.~\eqref{eq:H0}, it is instructive to recapitulate what is known about the analogous question for the SU(2) invariant quantum antiferromagnet. For the latter, neither triangular nor square lattices display a spin-liquid ground state (but rather 120$^\circ$ collinear and N\'eel antiferromagnetism), despite the fact that the mean field Abrikosov-fermion treatment yields Dirac spin-liquids.\,\cite{AffleckMarston1988,KotliarLiu1988,IqbalBecca2016}
In the present, predominantly ferromagnetic case, the situation is likely similar, and indeed the in-plane ferromagnetic solution with energy $-3J$ ($-2J$) is lower than any of the tRVB trial states of table~\ref{tab:Meanfieldsolutions} in a substantial fraction of the parameter regime. Therefore, it is an important task for the future to find an easy plane ($U(1)$ spin symmetric) Hamiltonian (potentially including longer-range and multi-spin interactions) which rigorously displays a tRVB ground state.

Regarding the concrete material 1T-TaS$_2$ (and related monolayer TaSe$_2$, NbSe$_2$), reliable understanding about the scale and sign of the exchange interactions is necessary, both experimentally and theoretically. Here we have presented a theory which displays quantum spin-liquid behavior, even when the exchange interactions are predominantly ferromagnetic. Incidentally, within our theory, doping leads to a time reversal symmetry breaking superconductor which is indeed {believed to be} observed in stacked multilayers of 1T-TaS$_2$ and metallic 1H-TaS$_2$.\,\cite{RibakKanigel2020} On the other hand, if present estimates of exchange interactions on the order of a few Kelvin are correct, an alternative to the theoretical QSL explanation for the weak Curie-Weiss signal appears inevitable. 

{Beyond the issue of 1T-TaS$_2$, our proposal opens room for a new and exciting experimental and numerical search for QSLs and topological order in local moment systems with predominantly ferromagnetic coupling that do not order down to very low temperatures.}

\section{Acknowledgments}

It is a pleasure to acknowledge useful discussions with L. Classen, B. J\"ack and D. Pasquier.

This work was supported by the U.S. Department of Energy, Office of Basic Energy Sciences, under Contract No.
DE-FG02-99ER45790 (E.J.K., P.C.), the National Science Foundation Grant No. DMR-1830707 (Y.K.) and was completed at the Aspen Center for Physics,
which is supported by National Science Foundation grant PHY-1607611.

\section*{APPENDIX}

\appendix

\section{Mean field slave boson theory for SU(2)}
\label{app:DetailsSpinon}

In this appendix we provide details on our calculations of a $t-J$ model with anisotropic, ferromagnetic exchange interaction. 

%

\subsection{Attractive interaction channels}

The spinon interaction in Eq.~\eqref{eq:HSB} contains intersite interaction of singlet particle-hole $f^\dagger_i f_j$ and particle-particle $f^\dagger_i \sigma_y f_j^\dagger$ and of triplet $m_z = 0$ particle-hole $f^\dagger_i  \sigma_z f_j$ and particle-particle $f^\dagger_i \sigma_x f_j^\dagger$ operators. We drop the interaction of repulsive interaction channels of triplet $m_z = \pm 1$ operators and obtain
\begin{align}
H_J  =\sum_{\langle i,j \rangle} & \frac{1}{2}\Big [E_{\rm sVB} (f^\dagger_i f_j)(f^\dagger_j f_i)  \notag \\
&+E_{\rm tVB} (f^\dagger_i\sigma_z f_j)(f^\dagger_j \sigma_z f_i) \notag \\
&+E_{\rm tVB} (f^\dagger_i\sigma_x f_j^\dagger)(f_j \sigma_x f_i) \notag \\
&+E_{\rm sVB} (f^\dagger_i\sigma_y f_j^\dagger)(f_j \sigma_y f_i)\Big ].
\end{align}
Since we are interested in the regime $E_{\rm tVB} < E_{\rm sVB}$ and $E_{\rm tVB}<0$, we decouple only the second and third line in Eq.~\eqref{eq:Decoupled} of the main text.

\subsection{Free energy expansion}
\label{app:FreeEnergyExpansion}

We expand the fermionic contribution to the free energy density, Eq.~\eqref{eq:FreeEnergy}, in powers of $\kappa', \vec u$.

The zeroth order is given by 
\begin{equation}
F_0 = 2\frac{T}{\mathcal A} \sum_{\v k} \ln (1 + e^{- \xi_{\v k}/T}),
\end{equation}
while higher order contributions are formally
\begin{align}
\Delta F &= - \frac{T}{2 \mathcal A} \sum_{n,\v k} \tr^{\sigma, \tau} \ln[\mathbf 1 - \mathcal G^{(0)}_{n,\v k} \delta h_{\v k}] \notag  \\
&\simeq  \frac{T}{2 \mathcal A} \sum_{n,\v k} \left ( \frac{1}{2} \tr^{\sigma, \tau} [(\mathcal G^{(0)}_{n,\v k}  \delta h_{\v k})^2 ] + \frac{1}{4} \tr^{\sigma, \tau} [(\mathcal G^{(0)}_{n,\v k}  \delta h_{\v k})^4 ]\right ).
\end{align}
Here $\mathcal G_{n, \v k} =  \left (  \begin{array}{cc}
G_{n, \v k} & 0 \\ 
0 & -\bar {G}_{n, \v k}
\end{array} \right )$ and $G_{n, \v k} = (i\epsilon_n - \xi_{\v k})^{-1}$. All terms with odd powers in the series expansion vanish by spin summation. This is in contrast to usual singlet RVB theory, where there is a linear term $\Delta F \sim -t_f \sum_l\kappa'_l$.

\subsubsection{Second order term}

We introduce the three integrals
\begin{subequations}
\begin{eqnarray}
A_\pm^{(ll')} &=& \frac{T}{{\mathcal A}} \sum_{n, \v k} \sin(k_l)\sin(k_{l'}) \left ( \frac{1}{i \epsilon_n - \xi_{\v k}} \pm \frac{1}{i \epsilon_n + \xi_{\v k}} \right)^2 \notag \\
&=& -\frac{1}{{\mathcal A}}\sum_{\v k}\sin(k_l)\sin(k_{l'}) \notag\\
&&\times\left \lbrace\frac{1}{2T[\cosh(\xi_{\v k}/[2T])]^2} \pm \frac{ \tanh(\xi_{\v k}/[2T])}{\xi_{\v k}} \right \rbrace,\\
B^{(ll')} &=& \frac{T}{\mathcal A} \sum_{n, \v k} \cos(k_l)\cos(k_{l'}) \left ( \frac{1}{(i \epsilon_n - \xi_{\v k})^2} + \frac{1}{(i \epsilon_n + \xi_{\v k})^2} \right) \notag  \\
&=& -\frac{1}{2T{\mathcal A}}\sum_{\v k}\frac{\cos(k_l)\cos(k_{l'})}{[\cosh(\xi_{\v k}/[2T])]^2}.
\end{eqnarray}
\label{eq:integrals1}
\end{subequations}

Using this integrals 
we obtain after traces in Nambu and spin space (Einstein summations to be understood)
\begin{eqnarray}
F_2 &=& A_+^{(ll')} \vec u_l \cdot \vec u_{l'} - A_-^{(ll')} \vec u_l ^T \left (\begin{array}{ccc}
1 & 0 & 0 \\ 
0 & 1 & 0 \\ 
0 & 0 & -1
\end{array}  \right) \vec u_{l'} \notag\\
&& +  2B^{(ll')}  \kappa_l' \kappa_{l'}'.
\end{eqnarray}

In Eq.~\eqref{eq:FreeEnergy} of the main text we slightly abuse our notations and absorb $A_+^{(ll')} + 2 \delta_{ll'}/\vert E_{\rm tVB}\vert \rightarrow A_+^{(ll')}$

\subsubsection{Mean field instabilities}

We first consider the regime $t_f = \delta t \ll T \ll t$. In this limit, $\partial f/\partial \lambda = 0$ leads to $\lambda \approx 2 \delta T \ll t_f$. Expansion in $\xi_{\v k}\ll T$ leads to
\begin{eqnarray}
A_{+}^{ll'} & \simeq & - \frac{1}{T \mathcal A} \sum_{\v k} \sin(k_l) \sin(k_l') \left (1 - \xi_{\v k} ^2/6\right) \notag \\
&\simeq & - \frac{\delta_{ll'}}{2T}\begin{cases} \left[ 1 - \frac{5}{6} \left (\frac{t_f}{T} \right)^2\right ], & \text{triangular latt.,} \\
\left[ 1 - \frac{3}{6} \left (\frac{t_f}{T} \right)^2\right ], & \text{square latt.,} \\
\end{cases} \\
A_{-}^{ll'} & \simeq &  \frac{1}{T \mathcal A} \sum_{\v k} \sin(k_l) \sin(k_l')  \xi_{\v k} ^2/12 \notag \\
&\simeq & \frac{\delta_{ll'}}{24 T} \begin{cases} \left[ 5 \left (\frac{t_f}{T} \right)^2\right ], & \text{triangular latt.,} \\
\left[ 3 \left (\frac{t_f}{T} \right)^2\right ], & \text{square latt.,} \\
\end{cases} \\
B &\simeq & - \frac{1}{2 T \mathcal A} \sum_{\v k} \cos(k_l) \cos(k_l') (1- \xi_{\v k}^2/4)\notag \\
&\simeq& - \frac{1}{4T}\begin{cases} \left[ 1 - \frac{1}{4}\left (\frac{t_f}{T} \right)^2 \left (\begin{array}{ccc}
7 & 2 & 2 \\ 
2 & 7 & 2 \\ 
2 & 2 & 7
\end{array}  \right)\right ], & \text{triangular latt.,} \\
\left[ 1 - \frac{1}{4} \left (\frac{t_f}{T} \right)^2 \left (\begin{array}{cc}
5 & 2 \\ 
2 & 5
\end{array}  \right) \right ], & \text{square latt..} \\
\end{cases}
\end{eqnarray}
The largest transition temperature $T_c$ is thus associated to the order parameter $\Delta_l', \Delta_l''$ with $T_c = \vert E_{\rm tVB}\vert/4 (1-  20(t_f/\vert E_{\rm tVB}\vert)^2/3)$ [$T_c =  \vert E_{\rm tVB}\vert/4(1  -5(t_f/\vert E_{\rm tVB}\vert)^2)$] for the triangular [square] lattice. The critical temperatures associated to $\kappa$ are order $(t_f/J)^2$ {times} lower.

We now consider the regime $T \ll t_f$. A logarithmic Cooper instability in Eq.~\eqref{eq:integrals1}, manifest through the typical momentum sum $\sum_{\v k} \tanh[\xi_{\v k}/(2T)]/\xi_{\v k}$, {only occurs for the combination $A_+^{(ll')} - A_-^{(ll')} \sim -\rho \delta_{ll'} \ln(t_f/T)$} (right at a van-Hove singularity, other terms may also have logarithmic coefficients).  
Here we employed $t_f$ as the UV cut-off of our theory and we introduced the density of states $\rho$ which in the simplified, parabolic limit obtained by expansion of the dispersion about the $\Gamma$ point is $\rho = 1/( 8 \sqrt{3}\pi t_f)$ [$\rho = 1/(8 \pi t_f)$] for triangular [square] lattice. Clearly, the parabolic approximation is considerably better for the triangular lattice which does not display a van-Hove singularity at half-filling.

We observe that $\kappa'$ does not develop a (dominant) mean field instability in either $T \ll t_f$ nor $t_f \ll T \ll t$. It will therefore be omitted in the following and is not included in Eq.~\eqref{eq:GL} of the main text.

\subsubsection{Fourth order term}

We obtain the following expansion of the trace
\begin{eqnarray}
\tr^{\sigma \tau} \left [\left (\mathcal G \vec h \cdot \vec \tau \sigma_z \right )^4 \right ]/2 &=& 2  \vert G \vert^4 \vec h^4  \notag \\
&-& 4  \vert G \vert^2(G+ {\bar G})^2  (h_1^2 + h_2^2)h_3^2\notag \\
&+& (G^2 - \bar G^2)^2 h_3^4 . \label{eq:FourthTerms}
%
%
\end{eqnarray}
Here, we have omitted the subscript $n, \v k$ of the Green's functions, and $\vec h = \vec h(\v k) = 2 \sum_l \vec u_l \sin(k_l)$.

We keep only the first line $2 \vert G \vert^4\vec h^4$ of Eq.~\eqref{eq:FourthTerms}. In the limit of small $t_f \ll T$ this is justified, since the terms of the second and third line vanish (this is a manifestation of the SU(2) symmetry). In the opposite limit, $t_f \gg T$ the leading, quadratic instability regards superconducting order parameters $\Delta'_l \Delta''_l$, only, and we can disregard the additional corrections from $h_3$ terms. 

In summary, this motivates us to drop SU(2) breaking terms from the quartic term.
The relevant integrals in the derivation of quartic terms are thus
\begin{equation}
I_{\lbrace k_i\rbrace } = \frac{{2} T}{\mathcal A}\sum_{n, \v k} \vert G_{n, \v k} \vert^4 \prod_{i = 1}^4 \sin(k_{l_i}).
\end{equation}
By mirror symmetry $k_1 \rightarrow - k_1$, the only non-zero integrals have either all $l_i$ equal, or $l_i$ pairwise equal. By rotational symmetry, all non-zero integrals can thus be expressed as one of the following integrals
\begin{subequations}
\begin{eqnarray}
\tilde C &=& \frac{{2} T}{\mathcal A}\sum_{n, \v k} \vert G_{n, \v k} \vert^4 \sin(k_1)^4 \notag \\
&=& \frac{7 \zeta(3)}{4\pi^2 T^3}\frac{1}{\mathcal A} \sum_{\v k} f(\xi_{\v k}/T)  \sin(k_1)^4, \\
D &=& \frac{{4} T}{\mathcal A}\sum_{n, \v k} \vert G_{n, \v k} \vert^4 \sin(k_1)^2\sin(k_2)^2  \notag \\
&=& \frac{7 \zeta(3)}{4\pi^2T^3}\frac{2}{\mathcal A} \sum_{\v k} f(\xi_{\v k}/T)  \sin(k_1)^2\sin(k_2)^2.
\end{eqnarray}
\label{eq:integrals2}
\end{subequations}
Here, $f(x)=-\pi^2{(x -\sinh (x)) \text{sech}^2\left({x}/{2}\right)}/(7 \zeta(3) {x^3})$ (which is normalized to $\int dx f(x)= 1$) places the integrals on the Fermi surface.

In the regimes of interest we estimate these constant as follows: for $t_f \ll T \ll t$, $C, D \sim T^{-3} \sim J^{-3}$, while for $T \ll t_f$, $C, D \sim T^{-2}/[t_f] \sim J^{-2}/[t_f]$.

With this notation we obtain
\begin{equation}
F_4 = \sum_{l} \tilde C \vec u_l^4 + \frac{D}{2} \sum_{l \neq l'} [\vec u_l^2 \vec u_{l'}^2 + 2(\vec u_l \cdot \vec u_{l'})^2 ]
\end{equation}

as reported in Eq.~\eqref{eq:GL} of the main text, where we use the notation $C = \tilde C - 3D/2$.


\section{Large N treatment of tRVB spin liquid}
\label{app:largeN}

In this appendix, we demonstrate that the triplet quantum spin liquid discussed in the main text can be stabilized in an appropriate large $N$ limit. Here we concentrate on the Mott insulator (i.e. $\delta =0$) and $J_z = J$. In this limit, Eq.~\eqref{eq:H0} can be written as
\begin{equation} \label{eq:H0App}
H = J \sum_{\langle i,j \rangle} \sum_{\mu = 1}^3 (\hat \sigma_z^{(i)} \hat \sigma_\mu^{(i)} \hat \sigma_z^{(i)}) \hat \sigma_\mu^{(j)}.
\end{equation}

The easy plane nature is captured by the unitary  transformation $\hat \sigma_z^{(i)} \hat \sigma_\mu^{(i)} \hat \sigma_z^{(i)}$ on the $i$-site. Note that Eq.~\eqref{eq:H0App} holds for triangular and square lattices, alike. 

Here we introduce a generalization of the $SU(2)=Sp(2)$ spin group to $Sp(2N)$ and the large-N model is 
\begin{equation} \label{eq:H0SpN}
H = \frac{J}{N/3+2/3} \sum_{\langle i,j \rangle} \sum_{\mu = 1}^{2N^2 + N} (\hat \sigma_z^{(i)} \hat \sigma_\mu^{(i)} \hat \sigma_z^{(i)}) \hat \sigma_\mu^{(j)}.
\end{equation}

The notion of $\hat \sigma_z$ operators on a given site $i$ will become clear shortly and the prefactor is chosen for convenience. 
The symplectic group is special, as it allows for the notion of time-reversal symmetry and thereby for superconducting spinon mean field theories.\,\cite{FlintDzeroColeman2008} This feature is manifest in the usual matrix definition of the generators of the group
\begin{equation} \label{eq:SpNDef}
\hat \sigma_\mu = - \hat \sigma_y \hat \sigma_\mu^T \hat \sigma_y,
\end{equation}
where $\hat \sigma_y = \left (\begin{array}{cc}
0 & -i \mathbf 1_N \\ 
i \mathbf 1_N & 0
\end{array} \right) \equiv \sigma_y \otimes \mathbf 1_N$ (where $\mathbf 1_N$ is the $N\times N$ identity matrix).

For the present purpose of anisotropic quantum magnetism, it is convenient to represent the generators of $Sp(2N)$ as $\sigma_{x,y,z} \otimes S$ and $\mathbf 1_2 \otimes A$, in particular $\hat \sigma_{y,z} = \sigma_{y,z} \otimes \mathbf 1_N$ in Eqs.~\eqref{eq:H0SpN},\eqref{eq:SpNDef}. Here, $S$ ($A$) are symmetric (antisymmetric) $N\times N$ matrices, clearly this parametrization fulfills Eq.~\eqref{eq:SpNDef} and the number of generators $3 N(N+1)/2 + N(N-1)/2 = 2N^2 +N$ agrees with the dimension of the group.
We can thus rewrite Eq.~\eqref{eq:H0SpN} as
\begin{align} \label{eq:H0SpNExpl}
H &= -\frac{J}{N/3+2/3} \sum_{\langle i,j \rangle} \Big \lbrace \sum_{S} [(\sigma_x \otimes S)^{(i)}(\sigma_x \otimes S)^{(j)}\notag \\
&+(\sigma_y \otimes S)^{(i)}(\sigma_y \otimes S)^{(j)}-(\sigma_z \otimes S)^{(i)}(\sigma_z \otimes S)^{(j)}] \notag \\
&- \sum_A (\mathbf 1_2 \otimes A)^{(i)}(\mathbf 1_2 \otimes A)^{(j)} \Big \rbrace.
\end{align}
This expression clearly reflects the easy plane ferromagnetism of the original XXZ spin-1/2 model Eq.~\eqref{eq:H0} of the main text (here $\sum_S$ sums over symmetric $N \times N$ matrices and analogously $\sum_A$ sums over antisymmetric matrices). 

We now represent the symplectic spins as
\begin{equation}
\hat \sigma_\mu \rightarrow f^\dagger_{a, \sigma} [\hat \sigma_\mu]_{\substack{a a'\\
\sigma \sigma'}} f_{a, \sigma}.
\end{equation}
Here, we introduced $2N$ species ($\sigma = \uparrow, \downarrow$, $\alpha = 1 \dots N$) of Abrikosov fermions. 

In this notation, Eq.~\eqref{eq:H0SpNExpl} becomes
\begin{eqnarray}
H &=& \frac{J}{N/3+2/3} \sum_{\langle i, j \rangle} \sum_{\mu = 1}^{2N^2 + N} (f^\dagger_i \hat\sigma_z \hat\sigma_\mu \hat\sigma_z f_i) (f^\dagger_j \hat\sigma_\mu f_j) \notag \\
&=& -\frac{J}{2N/3+4/3} \sum_{\langle i, j \rangle} (f^\dagger_i \hat\sigma_z f_j)( f_j^\dagger \hat\sigma_z f_i) \notag \\
&&-\frac{J}{2N/3+4/3} \sum_{\langle i, j \rangle} (f^\dagger_i \hat\sigma_z \hat\sigma_y f_j^*)( f_j^T \hat\sigma_y \hat\sigma_z f_i) 
\end{eqnarray}
In the second line we used the Fierz identity of generators of appropriately normalized generators of $Sp(2N)$ 
\begin{equation}
\sum_\mu \hat \sigma^\mu_{\alpha \beta} \sum_\mu \hat \sigma^\mu_{\gamma \delta}  = \frac{1}{2} \left ( \delta_{\alpha \delta} \delta_{\beta \gamma} - (\hat \sigma_y)_{\alpha\gamma }(\hat \sigma_y)_{ \delta \beta} \right).
\end{equation}
In this equation we used a multi-index notation $\alpha = (\sigma, a)$ etc.

Clearly, the interaction is invariant under local transformations which leave
$f_if_i^\dagger + \sigma_y f_i^* f_i^T \sigma_y$ invariant, which leads to a local SU(2) symmetry.\,\cite{FlintDzeroColeman2008} Here we will not dwell on these subtleties, and rather exploit that the large $N$ limit prescribes the pattern of decoupling that we employed in Eq.~\eqref{eq:Decoupled}
\begin{eqnarray}
H &=& \sum_{\langle i,j \rangle} \lbrace( \kappa_{ij} f^\dagger_i \hat\sigma_z f_j + H.c.) + [\Delta_{ij} f_i^T(i \hat \sigma_y) \hat\sigma_z f_j + H.c.]\rbrace\notag \\
&+& \frac{2 N}{\vert E_{\rm tVB}\vert } \sum_{\langle i,j \rangle}[ \vert \kappa_{ij} \vert^2 + \vert \Delta_{ij} \vert^2 ].
\end{eqnarray}
We used $E_{\rm tVB} = -3J$ in the present limit. Upon integration of the spinons, the overall free energy therefore acquires an additional linear proportionality to ``color index'' $N$, whereby the mean-field approximation becomes justified. 

Before closing, we remark that our search of mean-field solution within the limited set of homogeneous $\kappa_{i,j}, \Delta_{i,j}$ is not controlled by the large $N$ limit and deserves special attention in future studies of the problem. 

In summary, in this appendix we have presented an $Sp(2N)$ generalization of Eq.~\eqref{eq:H0} which allows to analytically control the mean field treatment of tRVB theories. It is apparent that this representation of $Sp(2N)$ is favorable for any anisotropic quantum magnet and the application to Kitaev-Heisenberg models is left to future publications.\,\cite{KoenigUnpublished}

\section{BKT transition}
\label{app:BKT}

Here we summarize the mapping of the bosonic part of the Eq.~\eqref{eq:Decoupled}
\begin{align}\label{eq:NNBOseHubbard}
H_{b} &= - t_b \sum_{\langle i,j \rangle}[b_i^\dagger b_j + H.c.] -\tilde \mu \sum_i  b_i^\dagger b_i + V\sum_{\langle i,j \rangle} (b_i^\dagger b_i)(b_j^\dagger b_j)
\end{align} 
to the continuum model, Eq.~\eqref{eq:ContBosons} and thereby we derive $T_{\rm BKT}$. {We assume a grand canonical ensemble and here use} $\tilde \mu = \mu -2zV - \lambda$ {to denote the effective boson chemical potential.} 
The {microscopic values for the} 
coupling constants of Eq.~\eqref{eq:ContBosons} 
is {derived in this section and} summarized in Tab.~\ref{tab:CoefficientsContTheory}.

In the weak coupling limit $t_b \gg V$ we can expand $\xi^b_{\v k} = -2 t_b \sum_l \cos(\v k \cdot \hat e_l) \simeq (-z + \frac{z \v k^2}{2}) t_b$ near the bottom of the band and thus identify $\xi^b_{0} = -zt_b$ (which is absorbed in $\mu$) and the mass $m = 1/(zt_b a^2)$. Then, Eq.~\eqref{eq:ContBosons} can be directly derived by replacing $b_i \rightarrow a \psi(\v x_i)$, {with $a$ being the lattice constant}.

For the regime is $t_b \ll V$, we expand in small order parameter field $\Psi$ by decoupling the kinetic term (for details see Ref.~\onlinecite{SachdevBook})
\begin{equation}
H_{\rm kin} = \sum_{\v k} b_{\v k}^\dagger E(\v k) b_{\v k} \rightarrow \sum_i b_i^\dagger \Psi_i + \bar \Psi_i b_i - \sum_k \bar \Psi(\v k) E(\v k)^{-1} \Psi(\v k).\label{eq:BoseKin}
\end{equation}

{In Eq.~\eqref{eq:BoseKin}}, the term linear in $\Psi$ will be treated as a perturbation to $H_{b}\vert_{t_b = 0}$, in which local occupations $n_i$ are good quantum numbers allowing to formally diagonalize the Hamiltonian. In particular, the state without any boson has energy $0$, while $E_1 = -  \tilde \mu $ is the energy of a single boson on a given site, while $E_V = 2 E_1 + V$ is the energy of a pair of bosons at adjacent sites.

The continuum theory of bosons, Eq.~\eqref{eq:ContBosons}, near $\mu_{\rm eff} = 0$ describes a (Mott-) insulator to superfluid quantum phase transition, and we formally derive this theory approaching the transition from the disordered side with $\mu_{\rm eff} <0$ such that no bosons are in the system at the ground state. We then assume that the parameters $m, g, \mu_{\rm eff}$ are the same also on the ordered side of the transition (in the relevant regime $\mu_{\rm eff} \sim \delta g/a^2$ is small but positive) and at temperatures of order $T_{\rm BKT}$.

By assumption, the ground state of $H_{b}\vert_{t_b = 0}$ is then given by all sites {empty} and the leading excitations are given by the 1- and 2- boson states discussed above. 
We obtain from integrating {out} the $b_i$-bosons\,\cite{SahaPixley2020}
\begin{eqnarray}
S &=& \int d\tau \sum_i \frac{\bar \Psi_i (-1 +E_1 \partial_\tau) \Psi_i}{E_1} \notag \\
&+& 4 \int d\tau \sum_{\langle i, j \rangle}\frac{\vert \Psi_i \vert^2\vert \Psi_j \vert^2}{E_1^2} \left (\frac{1}{2 E_1}-\frac{1}{E_V} \right).
\end{eqnarray}

The quadratic term is obtained by perturbative inclusion of the dynamics of a single boson in the system. The quartic term can be obtained in the same manner, but it is equivalent and easier to calculate perturbative corrections to the ground state energy of $H_{b}\vert_{t_b = 0}$ in the presence of a constant $\Psi$. 
The parameters of Tab.~\ref{tab:CoefficientsContTheory}, third column,
{and Eq.~\eqref{eq:ContBosons} are} then obtained by identifying $\Psi_i \rightarrow  a \psi(\v x_i) E_1$ for weakly spatially dependent fields.

{Physically, the presented calculation can be interpreted as follows. First, consider creating a single boson in the system. It costs onsite energy $E_1= -\tilde \mu>0$, but gains kinetic energy up to $zt_b$ as it travels through the system. Clearly, the physics of a single boson is approximately valid in the small density limit, too, thus the free part of Eq.~\eqref{eq:ContBosons} is equivalent for non-interacting, weakly interacting and strongly interacting particles. However, the free theory needs to be complemented by scattering between two bosonic waves when two bosons are close (i.e. the $g$ term). This depends on the microscopic details of the interaction potential, in particular its strength, as presented in Fig.~\ref{fig:mg}.}

\begin{table}
\begin{tabular}{|c||c|c|}
\hline
 & $t_b \gg V$ & $t_b \ll V$ \\
\hline \hline
$m$ & $1/(z t_b a^2)$ & $1/(z t_b a^2)$\\
\hline
$\mu_{\rm eff}$ & $\mu-2 zV - \lambda + zt_b$ & $\mu-2 zV - \lambda + z t_b$ \\
\hline
$g$ & $2zV a^2$ & $4z^2 t_b a^2(1- {2z t_b}/{ V})$ \\
\hline
$T_{\rm BKT}$ & $\frac{2\pi z t_b n a^2}{\ln\left ( {190 t_b}/{V}\right)}$ &$\frac{2\pi z t_b n a^2}{\ln\left (95/(z - 2 z^2 t_b/V)\right)}$\\
\hline
\end{tabular}
\caption{Coupling constants of the continuum field theory of 2D bosons Eq.~\eqref{eq:ContBosons} as obtained from the bosonic part of Eq.~\eqref{eq:Decoupled} (here, we explicitly restored the lattice constant $a$).}
\label{tab:CoefficientsContTheory}
\end{table}

\bibliography{tRVBKotliarLiu}

\begin{thebibliography}{64}
\bibitem [{Foo()}]{FootnotePasquierYazyev2018}%
  \BibitemOpen
  \href@noop {} {}\bibinfo {note} {The ferromagnetic exchange interaction of $J
  \sim 5 K$ has been derived for monolayer 1T-NbSe$_2$. We are not aware of a
  similar derivation for bulk 1T-TaS$_2$, but expect a similar number from
  analogous techniques.}\BibitemShut {Stop}%
\end{thebibliography}

 \end{document}